\documentclass[10pt]{article}

\usepackage{amsmath}
\usepackage{amsfonts}
\usepackage{amssymb, multirow}
\usepackage{epsfig}
\usepackage{graphicx}
\usepackage{subfigure}
\usepackage{cite}

\begin{document}

\newcommand{\be}{\begin{equation}}\newcommand{\ee}{\end{equation}}
\newcommand{\bea}{\begin{eqnarray}} \newcommand{\eea}{\end{eqnarray}}
\newcommand{\ba}[1]{\begin{array}{#1}} \newcommand{\ea}{\end{array}}

\numberwithin{equation}{section}

\def\p{\partial }

\def\a{\alpha}
\def\b{\beta}
\def\m{\mu}
\def\n{\nu}
\def\r{\rho}
\def\s{\sigma}
\def\d{\delta}
\def\wg{\wedge}
\def\eps{\epsilon}
\def\veps{\varepsilon}
\def\nn{\nonumber}
\def\ov{\over }
\def\td{\tilde }

\newcommand{\var}{\varepsilon}

\newcommand{\cM}{{\cal M}}
\newcommand{\cN}{{\cal N}}
\newcommand{\cG}{{\cal G}}
\newcommand{\cL}{{\cal L}}
\newcommand{\cV}{{\cal V}}
\newcommand{\cK}{{\cal K}}
\newcommand{\cW}{{\cal W}}
\newcommand{\mZ}{\mathbb{Z}}
\newcommand{\mI}{\mathbb{I}}
\newcommand{\mR}{\mathbb{R}}
\newcommand{\nH}{n_{\rm H}}
\newcommand{\nV}{n_{\rm V}}
\newcommand{\bP}{\mathbb P}
\newcommand{\bQ}{\mathbb Q}
\newcommand{\eins}{\mbox{$1 \hspace{-1.0mm} \text{l}$}}

\newcommand{\ft}[2]{{\textstyle\frac{#1}{#2}}}
\newcommand {\Rbar} {{\mbox{\rm $\mbox{I}\!\mbox{R}$}}}
\newcommand {\Hbar} {{\mbox{\rm $\mbox{I}\!\mbox{H}$}}}
\newcommand {\Cbar}
            {\mathord{\setlength{\unitlength}{1em}
             \begin{picture}(0.6,0.7)(-0.1,0)
                \put(-0.1,0){\rm C}
                \thicklines
                \put(0.2,0.05){\line(0,1){0.55}}
             \end {picture}}}

\newcommand{\eqn}[1]{(\ref{#1})}

\newenvironment{matr}[1]
{\left[ \begin{array}{{#1}}}{\end{array} \right]}

\newcommand{\tw}{{\bf 20_{\rm S}}}
\newcommand{\tww}{{\bf 20_{\rm A}}}

\def\p{\partial}
\def\bfone{\relax{\rm 1\kern-.35em 1}}
\def\dop{{\rm d}\hskip -1pt}

\renewcommand{\thesection}{\arabic{section}}
\renewcommand{\theequation}{\thesection.\arabic{equation}}
\rightline{ICCUB-11-135}

\rightline{April, 2011}

\bigskip

\begin{center}

{\Large\bf  
Holographic Superconductors from Gauged Supergravity
}
\bigskip
\bigskip

{\it \large Francesco Aprile\footnotemark[1], Diederik Roest\footnotemark[2]  and Jorge G. Russo\footnotemark[1]\footnotemark[3]}
\bigskip

{\it
1) Institute of Cosmos Sciences and Estructura i Constituents de la Materia\\
Facultat de F{\'\i}sica, Universitat de Barcelona\\
Av. Diagonal 647,  08028 Barcelona, Spain\\
\smallskip
2) Centre for Theoretical Physics,\\
University of Groningen, \\
Nijenborgh 4, 9747 AG Groningen, The Netherlands\\
\smallskip
3) Instituci\'o Catalana de Recerca i Estudis Avan\c cats (ICREA)\\
Pg. Lluis Companys, 23, 08010 Barcelona, Spain\\
}
\bigskip
\bigskip

\end{center}
\bigskip

\begin{abstract}

\noindent
We consider minimal setups arising from
different truncations of ${\cal N}=8$  five-dimensional  $SO(6)$ gauged supergravity 
to study phase transitions involving spontaneous breaking of any of the $U(1)$ symmetries
in $U(1)\times U(1)\times U(1)\subset SO(6)$.
These truncations only keep the three relevant vector fields, four complex scalar fields carrying $U(1)$ charges, 
plus two neutral scalar fields required by consistency.
By considering thermal ensembles with different fixed $U(1)$ charge densities and solving 
the complete equations including the full back-reaction, in some cases we find  instabilities towards the formation
of hairy black holes, which lead to second order transitions, resulting from a thermodynamical competition between different sectors.
 We argue that these should be the dominant thermodynamical instabilities in the full ten-dimensional type IIB theory.
 In other cases we find unstable branches of hairy black holes that extend to temperatures above a critical temperature
(``retrograde condensation").
 The results can be used as a first step to understand new
 aspects of the phase diagram of large $N$  ${\cal N}=4$ $SU(N)$ super  Yang-Mills  theory with fixed charge densities.

\end{abstract}

\clearpage

\tableofcontents

\section{Introduction}

One of the important problems in holographic superconductivity \cite{Gubser:2008px,Hartnoll:2008kx}\footnote{For reviews and a more complete list of references, see \cite{Herzog:2009xv,Horowitz:2010gk}.}  is to disclose the precise dictionary between gravity and the condensed matter system. This requires what is called a ``top-down" approach, starting with ten-dimensional string theory or M-theory and a brane construction so that the field theory undergoing the phase transition can be explicitly understood \cite{Denef:2009tp,Gubser:2009qm,Gauntlett:2009dn,Gauntlett:2009bh,Donos:2010ax}. 
These theories contain many degrees of freedom whose dual operator may condense and break a $U(1)$ symmetry, thus leading to superconductivity.
Identifying the most relevant mode, the one which is dual to the operator that condenses first and dominates the thermodynamics, is a priori a complicated task. However, an important clue  lies in the 
empirical observation  that operators with lower dimensions and higher R-charges generically have a higher critical temperature, and hence are prone to condense earlier as  the temperature is lowered (see e.g.~\cite{Denef:2009tp,Gubser:2009qm, Aprile:2010ge}). Therefore, a first step in the top-down approach would be the construction of consistent truncations that include the  lightest modes in the mass spectra arising from the higher dimensional theory. 
Here we will consider the spectra of theories with maximal supersymmetry that are relevant for holography, where the lightest
modes are scalar fields, which can have negative masses without inducing perturbative instabilities in AdS.

The three maximal supergravities with Anti-de Sitter vacua, in $D=7,5$ and $4$, share a number of  features. First of all, they all come from Freund-Rubin reductions of ten- or eleven-dimensional parent theories over an $S^4, S^5$ and $S^7$, respectively. As a consequence of the $S^{n-1}$ reduction, the gauge group is always $SO(n)$. This is embedded in an $SL(n)$ subgroup of the full global symmetry group, which is $SL(5)$, $E_6$ and $E_7$, respectively. 
In view of the above, it is important to elucidate the mass specta of the scalar fields of these theories.

In the case of AdS backgrounds, a lower bound on scalar masses is set by the Breitenlohner-Freedman (BF) bound at \cite{Breitenlohner:1982bm}
 \begin{align}
  m_{\rm L}^2 L^2 = - \tfrac14 (D-1)^2 \,,
 \end{align} 
where the scale $L$ of AdS is set by the scalar potential via the relation $V L^2 = -(D-1)(D-2)$, with the scalar potential evaluated in a critical point.
In addition to the above, there are two other interesting values of the masses slightly above the BF bound \cite{Townsend:1984iu}. The first is at 
 \begin{align}
  m_{\rm C}^2 L^2  = - \tfrac14 (D-1)^2 + \tfrac14 \,, 
 \end{align}
and corresponds to a conformally coupled scalar field. Finally, there is
 \begin{align}
  m_{\rm U}^2 L^2 = -\tfrac14 (D-1)^2 + 1 \,.
 \end{align}
This is the upper bound for scalar masses that still allow for two different boundary conditions. The dimension $\Delta$ of the dual operator is related by $m^2 L^2 = \Delta (\Delta - D + 1)$. The unitarity bound implies that the lowest possible dimension is $\Delta = \tfrac12 (D-3)$, corresponding to the upper bound on the mass. Upon increasing the dimension, the corresponding mass first goes down to the Breitenlohner-Freedman lower bound, after which it goes up to plus infinity. Both the masses and dimensions are illustrated in figure 1.

\begin{equation}
\begin{picture}(500,50)(-40,5)

\put(-40,20){$m^2 L^2 =$}
\put(-25,5){$\Delta=$}

\put(20,60){\vector(1,0){350}}
\put(20,60.3){\line(1,0){240}}
\put(20,59.7){\line(1,0){240}}

\put(20,60){\circle*{3}}
\put(17,40){L}
\put(0,20){$-\tfrac14 (D-1)^2,$}
\put(0,5){$\tfrac12 (D-1),$}

\put(80,60){\circle*{3}}
\put(77,40){C}
\put(60,20){$-\tfrac14 (D-1)^2+\tfrac14,$}
\put(60,5){$\tfrac12 (D-1 \pm 1),$}

\put(260,60){\circle*{3}}
\put(257,40){U}
\put(240,20){$-\tfrac14 (D-1)^2+1.$}
\put(240,5){$\tfrac12 (D-1 \pm 2).$}

\put(-40,-15){{\it Figure 1: the different values of allowed (negative) masses and dimensions in Anti-de Sitter. Masses on}}
\put(-40,-26){\it  the bold line allow for two physically relevant dimensions.}

\end{picture} \notag
\end{equation}

\vspace{1cm}

All three theories have an Anti-de Sitter vacuum in the origin, which is maximally supersymmetric and preserves the $SO(n)$ gauge group. On account of this, the scalars are organised in $SO(n)$ irreps with particular masses. It turns out that the three theories under consideration exactly have masses lying at the three special mass values listed above \cite{Townsend:1984iu}. In particular, the seven-dimensional theory has a 14-dimensional irrep of scalars with 
 \begin{align}
   D=7: \qquad m^2 L^2 = - 8 \,,
 \end{align}
corresponding to the upper bound. In contrast, the four-dimensional theory has two 35-dimensional irreps of scalars with  
\begin{align}
   D=4: \qquad m^2 L^2 = - 2 \,,
 \end{align} 
corresponding to a conformally coupled scalar field. Finally, the five-dimensional theory has 
 \bea
    D=5: \qquad && m^2 L^2 = - 4 \, \qquad 20\ {\rm scalars} \,,
 \\
 && m^2 L^2 = - 3 \, \qquad 20\ {\rm scalars} \,,
 \\
 && m^2 L^2 =  0 \, \qquad \ \ 2\ {\rm scalars} \,.
 \eea
 Thus 20 scalars saturate the BF bound, whereas 
another set of 20 scalars saturate the unitarity bound (more details will follow in the next section).
Therefore, the two maximal supergravities coming from M-theory have masses that satisfy the Breitenlohner-Freedman bound, while the $D=5$ theory coming from IIB has a number of scalars that saturate the bound. 

The  connection between type IIB string theory on AdS$_5$ $\times$ S$^5$ to large-$N$  ${\cal N}=4$ $SU(N)$ super Yang-Mills theory is the best understood AdS/CFT duality. In view of this,  we will focus on the five-dimensional supergravity case.
The relevant five-dimensional description is given by ${\cal N}=8$ $SO(6)$ gauged supergravity \cite{Gunaydin:1984qu, Pernici:1985ju} and  truncations thereof, which will be here 
used as a basic setting to understand superconductivity (or, more precisely, superfluidity) in ${\cal N}=4$  super  Yang-Mills  theory. 
This theory has $SO(6)$ global R-symmetry, which contains $U(1)\times U(1)\times U(1)$ as its maximal Abelian subgroup. 
In this context, one can thus consider a canonical ensemble with fixed charge densities $\hat \rho_1, \  \hat \rho_2,  \ \hat \rho_3$ associated with each $U(1)$ group.
The problem is then to understand what is the phase diagram of the theory as the temperature is gradually lowered from high values.
On the gravity side, the  high  temperature thermodynamics is dominated by the STU black hole with  charges   proportional to $\hat \rho_1, \  \hat \rho_2,  \ \hat \rho_3$.
As the temperature gets lower than some critical value, some scalar operators are expected to condense, possibly breaking $U(1)$ symmetries.
Understanding this condensation process in the full type IIB superstring theory context is obviously complicated.
However, the operators which should condense first are those with low dimensions, and hence are dual to light modes of $D=5$  ${\cal N}=8$ $SO(6)$ gauged supergravity. We will investigate a number of such modes for various configurations of black hole charges.

This paper is organised as follows. In the next section we will discuss maximal supergravity in five dimensions, together with a truncation to minimal supergravity and two non-supersymmetric truncations (sectors I and II). All of these theories have a $U(1)^3$ gauge group and a number of scalar
fields. In section 3 we will consider charged black hole solutions of this theory. In particular, we will first review the black hole solutions corresponding to the uncondensed phase. Subsequently the possibility of black holes with scalar hair, corresponding to a condensed phase, is investigated in the two non-supersymmetric sectors I and II. We will show which phase is thermodynamically dominant as a function of temperature, and relate our results to previous literature. In section 4 we discuss possible condensation from other sectors not included in the previous truncations.
Finally, some concluding remarks are given in section 5.

\medskip

\noindent {\it Note added:} 
We have been informed that a  complementary study of black hole solutions with hair of
$D=4$ gauged supergravity will appear in \cite{Donos:2011ut}, in concert with this paper.
A study of models related to ${\cal N}=4$ SYM with chemical potentials from the field theory side will appear in \cite{Huijse:2011hp}.

\section{Simple truncations of $D=5$ maximal gauged supergravity}
\setcounter{equation}{0}
 
\subsection{A supersymmetric truncation}

Maximal supergravity in $D=5$ has a global $E_6$ symmetry \cite{Cremmer:1980gs}. In particular, the scalar fields span the coset $E_6 / Usp(8)$. However, the presence of an $SO(6)$ gauging breaks this symmetry. Instead, it is convenient to arrange the fields in irreps of the $SL(2) \times SL(6)$ maximal subgroup or its compact subgroup. 

In terms of the $SL(2) \times SL(6)$
maximal subgroup, the isometries of $E_6$ decompose as
 \begin{align}
  \text{scalar isometries:~} ({\bf 1}, {\bf 35}) \oplus ({\bf 2}, {\bf 20}) \oplus ({\bf 3},{\bf 1}) \,.
 \end{align}
The first and last term correspond to $SL(6)$ and $SL(2)$, respectively, while 
the additional irrep in the middle combines these into $E_6$. Physical scalars correspond to the non-compact isometries, of which there are (of course) 42. In terms of $SO(2) \times SO(6)$, these are given by
 \begin{align}
 \text{scalars:~} ({\bf 1}, \tw) \oplus ({\bf 2}, \tww)^+ \oplus ({\bf 2},{\bf 1}) \,,
 \end{align}
where the $\tw$ denotes a symmetric rank-2 and the $\tww$ is an anti-symmetric rank-3 tensor. Moreover, the $+$ corresponds to an imaginary self-dual condition\footnote{In more detail, the $({\bf 2}, {\bf 20})$ transforms as a doublet of anti-symmetric three-forms of $SL(2) \times SO(6)$. Branching this to $SO(2) \times SO(6)$ one obtains two irreps: the (anti-)imaginary self-dual combinations  $({\bf 2}, \tww)^\pm$. One of these correspond to the compact generators of $E_6$, and hence is modded out, while the other combination is retained and corresponds to physical scalars.}. Surprisingly, the supermultiplet structure in $D=5$ is such that not all scalars have identical masses. Instead, the three irreps listed above have 
 \begin{align}
  m^2 L^2 = -4 \,, \;\;\; -3 \,, \;\;\;  0 \,,
 \end{align}
respectively \cite{Kim:1985ez}. The first two of these exactly correspond to the lower and upper bounds defined in the introduction.

The vector bosons are in the anti-symmetric representation of $SL(6)$:
 \begin{align}
  \text{vectors:~} ({\bf 1}, {\bf 15}) \,,
 \end{align}
which are of course massless. The theory also contains twelve two-forms:
 \begin{align}
 \text{two-forms:~} ({\bf 2},{\bf 6}) \,.
\end{align}
In the ungauged theory, the two-forms can be dualised to vectors and combine to form the $\bf 27$ irrep of $E_6$; however, due to the $SO(6)$ gauging this is no longer possible in the gauged supergravity. Instead, the two-forms acquire a mass term and have $m^2 L^2 = + 1$ \cite{Kim:1985ez}.

As for the fermions, these consists of 4 gravitini and 24 dilatini. The former are in the spinorial representation of $SO(6)$ and  have masses $m^2 L^2 = + \tfrac32$. The latter are in $\bf 4$ and $\bf 20$ representations with masses $- \tfrac32$ and $- \tfrac12$, respectively.

We now consider a consistent truncation of the maximal $SO(6)$ supergravity 
by modding out by a $\mathbb{Z}_2 \times \mathbb{Z}_2$ symmetry with generators\footnote{
This truncation was studied in \cite{Khavaev:2000gb} and more recently in \cite{Bobev:2010de}.} 
 \begin{align}
  \left( \begin{array}{cc} - \mathbb{I}_4 & \\ & +\mathbb{I}_2 \end{array} \right) \,, \qquad \left( \begin{array}{cc} + \mathbb{I}_2 & \\ & -\mathbb{I}_4 \end{array} \right) \,,
\label{zzz}
 \end{align}
acting on the fundamental representation of $SL(6)$. In fact, this corresponds to a supersymmetric truncation, as can be seen from the following. The $SO(6)$ transformations translate into the following $SU(4)$ transformation:
 \begin{align}
  \left( \begin{array}{cccc} +1 &&& \\  & -1 && \\ && +1 & \\ &&& -1 \end{array} \right) \,, \qquad \left( \begin{array}{cccc} +1 &&& \\  & -1 && \\ && -1 & \\ &&& +1  \end{array} \right) \,,
 \label{SU4}
 \end{align}
where we have related the fundamental of $SO(6)$ to the anti-symmetric representation of $SU(4)$ via invariant tensors, which are a permutation of the 't Hooft symbols. From \eqref{SU4} it follows that this truncation preserves one out of the four original supersymmetries (thus there are eight preserved 
supercharges). 

The resulting field content is as follows. Firstly, all two-forms are removed by the truncation. 
Secondly, only three vector bosons survive, associated with the maximal Abelian subgroup $U(1)\times U(1)\times U(1)$ in $SO(6)$. 
Finally, out of the three irreps of scalar isometries, we retain respectively eleven, sixteen and three isometries. The corresponding numbers of physical scalars are eight, eight and two. Similarly, on the fermionic side we find a single gravitino and six spin-1/2 fields. These fields get organized in the following multiplets of minimal supersymmetry in $D=5$:
 \begin{itemize}
  \item Gravity: the graviton, a gravitino and a vector,
  \item Vector: a vector, a gaugino and a real scalar,
  \item Hyper: a hyperino and four real scalars.
 \end{itemize}
Therefore the resulting $\cN = 1$ theory contains, in addition to the gravity multiplet, two vector and four hyper multiplets. The scalar manifold is given by
 \begin{align}
  SO(1,1)^2 \times \frac{SO(4,4)}{SO(4) \times SO(4)} \,,
 \end{align}
where the first factor is the very special Kahler geometry spanned by the vector multiplet, and the second factor is the Quaternionic-Kahler manifold of the hyper sector. The same result was found in \cite{Bobev:2010de}.

In what follows we will consider two different non-supersymmetric truncations, that share the two dilatons of the vector multiplets but pick out two completely orthogonal $(SL(2)/SO(2))^4$ scalar submanifolds of the hyper sector. In both truncations we will pick out the scalar mode(s) that are relevant to describe hair in various black hole backgrounds.


\subsection{Non-supersymmetric truncation I: Keeping scalars in the $\tw$}

We now make the first subsequent truncation, to bring the theory to a more manageable form. In addition to the $\mathbb{Z}_2 \times \mathbb{Z}_2$ generators acting on the fundamental of $SL(6)$, we also mod out by the $SO(2)$ generator $- \mathbb{I}_2$ acting on the fundamental of $SL(2)$. The latter has the following two implications:
 \begin{itemize}
 \item Supersymmetry is now lost: the corresponding transformation in terms of $U(4)$ leaves none of the four supersymmetries invariant.
 
 \item 
The $({\bf 2},\tww)$ irrep is removed by the truncation  and one is only left with the $SL(2) \times SL(6)$ scalar manifold.
 \end{itemize}
This corresponds to the five-sphere reduction of IIB supergravity with only the ten-dimensional metric and five-form retained\footnote{This truncation was previously studied in \cite{Liu:2007rv}.}.
Due to the second point, the scalar potential reduces to a universal formula, that is also valid for the $SL(n) / SO(n)$ subgroup of the other AdS maximal supergravities.
For the vectors and the $SL(n)$ subgroup of scalars, which we will parametrise by a symmetric matrix $T$, the Lagrangian reads\footnote{In addition there can be topological Wess-Zumino terms for the vectors fields. We have not included these as these will not play any role in what follows.} \cite{Cvetic:1999xx, Bergshoeff:2004nq}
 \begin{align}
  \cL = \sqrt{-g}\, [R - \tfrac14 {\rm Tr}[T^{-1} D_\mu TT^{-1}D^\mu T] - \tfrac14 {\rm Tr}[T^{-1} F_{\mu \nu} T^{-1} F^{\mu \nu}] - V] \,,
 \end{align}
where the covariant derivatives on the scalars read
 \begin{align}
  D_\mu T = \partial_\mu T + g_0 (A_\mu T - T A_\mu) \,.
 \end{align}
Finally, the scalar potential is given by
 \begin{align}
  V =  \tfrac{1}{2}\, g_0^2 \,  ( 2 {\rm Tr}[TT] - {\rm Tr}[T]^2) \,.
 \label{potencial}
\end{align}
The coupling constant $g_0$ sets the scale for the AdS radius $L$. In $D=5$ this relation reads $g_0 L = 1$. An important feature of this truncation in five dimensions is that it retains the $SL(6) / SO(n)$ scalar fields in the scalar potential with masses at the Breitenlohner-Freedman bound, which have a high chance to condense due to the low dimension of their dual operators. Finally, note that the additional $SL(2) / SO(2)$ scalars do not appear in these expressions and are massless and neutral. 

The combined action of the $SO(6)$  transformations \eqref{zzz} with the $SO(2)$ element $- \mathbb{I}_2$ leads to a field content containing three vectors and ten scalars. The latter span the non-compact part of the remaining symmetry group
 \begin{align}
  SO(1,1)^2 \times SL(2)^4 \,.
 \label{sm}
 \end{align}

The part of the $SL(6)$ scalar manifold that is invariant under the $\mathbb{Z}_2 \times \mathbb{Z}_2$ symmetries \eqref{zzz} can be parametrised by
 \begin{align}
  T_{mn} = \left( \begin{array}{ccc} X_1 \cM_1 &   & \\    &  X_2 \cM_2 &  \\  &   &  X_3 \cM_3 \end{array} \right) \,,
 \end{align}
where 
\begin{align}
  \cM_i =  \left( \begin{array}{cc} \cosh\eta_i + \sinh\eta_i \cos \theta_i & \sinh\eta_i\sin\theta_i \\ \sinh\eta_i\sin\theta_i & \cosh\eta_i - \sinh\eta_i \cos\theta_i\end{array} \right) \, , \label{SL2scalars}
 \end{align}
and $X_1 X_2 X_3 = 1$. A parametrisation of the latter is
 \begin{align}
   X_1 = e^{\varphi_1 -\varphi_2} \,, \quad
 X_2 = e^{\varphi_1  +\varphi_2 } \, ,\quad X_3 = e^{-2\varphi_1 } \, .
 \end{align}
We will use the same form \eqref{SL2scalars} for the separate $SL(2)$ scalar, parametrised by $\cM_4$. The four scalars are collected in $\eta_a = (\eta_i, \eta_4)$, and similar for $\theta_a$.

The Lagrangian of the remaining bosonic fields is then given by 
\begin{align}
\cL = & \sqrt{-g} \Big( R -   \sum_{i=1}^3 X_i{}^{-2} [\tfrac12 (\p X_i)^2  + \tfrac14  F^{i,\mu\nu} F^i_{\mu\nu} ]
-\tfrac12  \sum_{a=1}^4 [ (\p \eta_a)^2 +\sinh^2{\eta_a} \big(\p \theta_a +L^{-1} q_{ai} A_i\big)^2 ]
- V \Big) \,, \label{sectorI}
\end{align}
where we used the notation $A_i = (A^{12}, A^{34}, A^{56})$. These span the $U(1)^3$ remaining part of the gauge group. The scalar charges with respect to it are given by
 \begin{align}
  q_{1i}=(2,0,0)\ ,\quad q_{2i}=(0,2,0)\ ,\quad q_{3i}=(0,0,2)\ ,\quad q_{4i}=(0,0,0) \,.
 \end{align}
Finally, despite the absence of supersymmetry in this truncation, the resulting scalar potential can be written in terms of a superpotential and its derivatives with respect to $\eta_a$ and either $X_i$ or $\varphi_{1,2}$:
 \begin{align}
V & = \frac{1}{2L^2} \,  \left[\sum_{a=1}^4  \left( \partial W\over \partial \eta_a\right)^2 +  \sum_{i=1}^3 X_i^2 \left( {\cal D}_i W\right)^2\right]- 
\frac{1}{3L^2} \, W^2 \,, \notag \\
& = \frac{1}{2L^2} \,  \left[\sum_{a=1}^4  \left( \partial W\over \partial \eta_a\right)^2 +  \frac{1}{6}\left( {\partial W\over \partial \varphi_1} \right)^2 +  \frac{1}{2}\left( {\partial W\over \partial \varphi_2} \right)^2 \right]- 
\frac{1}{3L^2} \, W^2\ ,
 \end{align}
where
 \begin{align}
  W=  \sum_{a=1}^4 q_{ai} X_i  \cosh (\eta_a) \,,
\end{align}
and ${\cal D}_i W =  \partial_{i} W - \tfrac13 X_i^{-1} W$ represents a covariant derivative.

Note that the first three angular scalars $\theta_i$ are pure gauge and can be set equal to zero by a $U(1)^3$ transformation. It can be checked that the origin of moduli space is an extremum. Similarly, the eigenvalues of the Hessian of the scalar potential are given by
 \begin{align}
  m_a^2 L^2 = - \sum_{i=1}^3 q_{ai} q_{ai} \,. \label{Hessian}
 \end{align}
These are equal for the first three scalars $\eta_i$ and given by $- 4$. This indeed corresponds to all scalar masses saturating the Breitenlohner-Freedman bound. Finally, the last scalar $\eta_4$ drops out of the scalar potential on account of being neutral.

In what follows, three further truncations of this model will be considered. In particular, we will consider subsectors where a number of gauge vectors are identified, and focus on the dynamics of at most two scalar fields. As these truncations will later be used to study the emergence of hair in charged black holes, the number of gauge fields will determine the number of black hole charges. These will be used to label the various cases.

\subsubsection*{One single charge (I)}

First of all, we consider a truncation of the three-block model to a sector with just a single BH charge. 
In this case we augment the $\mathbb{Z}_2 \times \mathbb{Z}_2$ of \eqref{zzz} to a full $SO(4)$ that is orthogonal to the diagonal $SO(2)$.
In other words, we will now consider a  truncation of maximal supergravity, based on the decomposition of $SO(6)$ into $SO(4) \times SO(2)$. In terms of special unitary groups this reads $SU(4)$ into $SU(2) \times SU(2) \times U(1)$, where the two $SU(2)$ factors are block-diagonal generalisations of \eqref{SU4}. The different irreps of $SO(6) \simeq SU(4)$ then split up as
 \begin{align}
  {\bf 4} & \rightarrow {\bf 2}_{+1/2} \oplus {\bf 2}_{-1/2} \,, \notag \\
  {\bf 6} & \rightarrow {\bf 4}_0 \oplus {\bf 1}_{+1} \oplus {\bf 1}_{-1} \,, \notag \\
  {\bf 15} & \rightarrow {\bf 6}_0 \oplus {\bf 4}_{+1} \oplus {\bf 4}_{-1} \oplus {\bf 1}_0 \,, \notag \\
 \tw & \rightarrow {\bf 9}_0 \oplus {\bf 4}_{+1} \oplus {\bf 4}_{-1} \oplus {\bf 1}_{+2} \oplus {\bf 1}_{0} \oplus {\bf 1}_{-2} \,,
 \end{align}
in terms of $SO(4)$ irreps with an $SO(2)$ weight. Subsequently we only retain the singlets of $SO(4)$. This amounts to one vector, corresponding to the remaining $SO(2)\sim U(1)$ gauge group, and three scalar fields. The latter split up in one real  scalar, and one complex  scalar field with $U(1)$ charge $q=+2$. Furthermore, from the decomposition of the $\bf 4$ one can see that this truncation does not preserve any supersymmetry.

Instead of the diagonal $SO(2)$, we will use a different but equivalent embedding, where the $SO(2)$ acts on the last two indices. Consequenly, the $SO(4)$ acts on the first four indices. In terms of scalars, this truncation amounts to the Ansatz
 \begin{align}
   \eta_a = (0,0,\eta,0), \quad \theta_a = (0,0,\theta,0) \,, \quad \varphi_2 = 0 \,.
 \end{align}
Similarly, for the vector fields this requires $A_i = (0,0,A)$. The complete Lagrangian takes the form
\be
\cL= \sqrt{-g} \big[R  - 3 (\partial \varphi_1)^2- \tfrac12  (\partial \eta)^2 - \tfrac14 e^{4 \varphi_1} F_{\mu\nu}F^{\mu\nu}- \tfrac12 \sinh^2{\eta} ( \partial \theta + 2 L^{-1} A)^2
-  V  \big]\, ,
\label{dosnu}
\ee
with
\be
V = {1\over L^2}  e^{2\varphi_1}  \big ( -4 -2  e^{- 6 \varphi_1} +2 e^{- 6 \varphi_1}  \cosh^2\eta-8 e^{-3 \varphi_1} \cosh\eta\big) \,.
\label{tresnu}
\ee

\subsubsection*{Two equal charges (I)}

Similarly, one can consider a truncation to a subsector that can carry two of the three BH charges. In order to truncate to a smaller number of scalars, one can mod out by the $SO(6)$ element
\begin{align}
  \left( \begin{array}{cccc} & + \mathbb{I}_2 && \\ - \mathbb{I}_2 &&& \\ &&& + 1 \\ && -1 & \end{array} \right) \,.
 \end{align}
This has the following effect on the remaining scalars:
 \begin{align} 
  \eta_a = (\eta, \eta,0,0) / \sqrt{2} \,, \quad  \theta_a = (\theta, \theta, 0,0) \,, \quad \varphi_2 = 0 \,.
 \end{align}
Furthermore, consistent with the above truncation, we take the gauge vectors equal to
 \begin{align}
  A_i = (A,A,0) / \sqrt{2}  \,. 
 \end{align}
This implies that the Lagrangian reduces to
 \begin{align}
 \mathcal{L} = \sqrt{-g} \big[  R - 3 (\partial \varphi_1)^2 - \tfrac12 (\partial \eta)^2 - \tfrac14 e^{-2 \varphi_1  } F_{\mu \nu} F^{\mu \nu} -  \sinh^2(\eta/\sqrt{2}) (\partial \theta + \sqrt{2} L^{-1} A)^2 - V \big] \,,
\label{dosdi}
 \end{align}
with the scalar potential given by
 \begin{align}
  V = - \frac{4}{L^2} e^{- \varphi_1} ( 2 \cosh(\eta/\sqrt{2}) + e^{3 \varphi_1 }) \,.
 \label{tresdi}
\end{align}

\subsubsection*{Three equal charges (I)}

Finally, a very simple sector is obtained by further restricting to the `isotropic' truncation, where the three $SL(2)$ copies are identified with each other. Group-theoretically this can obtained by modding out by $SO(6)$ elements 
 \begin{align}
  \left( \begin{array}{ccc} & +\mathbb{I}_2 & \\ -\mathbb{I}_2  && \\ && \mathbb{I}_2 \end{array} \right) \,, \qquad \left( \begin{array}{ccc} + \mathbb{I}_2 && \\ && +\mathbb{I}_2 \\ & - \mathbb{I}_2 & \end{array} \right) \,.
 \end{align}
This requires the further restrictions on the scalar fields
 \begin{align}
  \eta_a = (\eta, \eta, \eta,0) / \sqrt{3} \,, \quad \theta_a = (\theta, \theta, \theta,0) \,, \quad 
  \varphi_1 = \varphi_2 = 0 \,.
 \end{align}
Similarly we set 
 \begin{align} 
  A_i = (A,A,A) / \sqrt{3} \,.
 \end{align} 
The resulting Lagrangian is
\be
  \cL = \sqrt{-g} \big[ R -\tfrac14 F^{\mu\nu}F_{\mu\nu}-
\tfrac{1}{2}  \p_\mu\eta \p^\mu \eta - \tfrac{3}{2} \sinh^2 ({\eta } / { \sqrt{3}}) (\partial \theta +{2} L^{-1} \,  A / \sqrt{3})^2 - V \big] \,,
\label{isot}
\ee
with
\be
V = - \frac{6}{L^2}\,   \left( 1+ \cosh^2({\eta / \sqrt{3}}) \right)
 \, .
\label{Visot}
\ee

\subsection{Non-supersymmetric truncation II: Keeping scalars in the $\tww$}

Next we consider a different truncation from the supersymmetric theory of section 2.1, where we retain $SL(2)$ scalars from the $({\bf 2}, \tww)^+$ rather than the $({\bf 1}, \tw)$ of $SO(2) \times SO(6)$. This truncation was considered in \cite{Khavaev:2000gb,Bobev:2010de}. It can be defined by keeping the invariant sector under a $\mathbb{Z}_4$  symmetry consisting of \eqref{zzz} augmented by the $SO(2) \times SO(6)$ element
 \begin{align}
  \left( \begin{array}{cc} & 1 \\ -1 & \end{array} \right) \otimes \left( \begin{array}{cccccc} & 1 &&&& \\ -1 &&&&& \\ &&& 1 && \\ && -1 &&& \\ &&&&& 1 \\ &&&& -1 & \end{array} \right) \,.
 \end{align}
Surprisingly, this truncation also leads to the scalar manifold \eqref{sm}. The origin of the $SO(1,1)^2$ scalars coincides with that of the previous section: these are still given by the scalars $\varphi_{1,2}$ and have masses $m^2 L^2 = -4$. In contrast, in this case the $SL(2)^4$ scalars come from the $({\bf 2}, \tww)^+$ and therefore have masses equal to $-3$. Therefore this truncation can be seen as being orthogonal to the one previously considered, at least in the scalar sector. As for the vectors, the above element retains the same three $U(1)$'s as in the previous two truncations.

The bosonic part of the resulting Lagrangian \cite{Bobev:2010de} formally looks like the bosonic Lagrangian \eqref{sectorI} of sector I modulo the following points:
 \begin{itemize}
 \item
 As pointed out before, the two $SO(1,1)$ dilatons $\varphi_{1,2}$ are common in both truncations, while the $(SL(2) / SO(2))^4$ scalars are completely orthogonal. For that reason we will refer to the charged scalars of sector II as $\td \eta_a$. 
 \item
 The charges of the four charged scalars w.r.t.~the $U(1)^3$ gauge group in this case are given by
 \begin{align}
  q_{1i}=(1,1,-1)\ ,\quad q_{2i}=(1,-1,1)\ ,\quad q_{3i}=(-1,1,1)\ ,\quad q_{4i}=(1,1,1) \,.
 \label{neoc}
\end{align}
Note that the Hessian of the scalar potential \eqref{Hessian} in this case indeed gives rise to masses at $-3$, corresponding to the upper bound on scalar masses that allow for two different boundary conditions. 
 \end{itemize}

Similar to the previous case, we will now consider the simplest Ans\" atze within this sector which are consistent with configurations of one single charge, two equal charges and three equal charges.

\subsubsection*{One single charge (II)}

(i) We consider configurations with a single charge, corresponding to $A_i = (0,0,A)$.
In this case it is consistent to set 
 \begin{align}
  \td \eta_a = (\eta,0,0,\eta) / \sqrt{2} \,, \quad 
 \td \theta_a = (-\theta,0,0,\theta) \,, \quad \varphi_2=0.
 \label{scalars}
 \end{align}
The Lagrangian takes the form
\be
  \cL = \sqrt{-g} \Big[ R - 3 (\p \varphi_1)^2  
-  \tfrac12 (\p \eta )^2 -  \sinh^2({\eta / \sqrt{2}}) \big(\p\td \theta +  L^{-1}  A \big)^2 - \tfrac14  e^{4\varphi_1} F_{\mu\nu} F^{\mu\nu}
- V \Big] \,,
\label{lsg}
\ee
where
\be
V= \frac{e^{-4\varphi_1 }}{L^2}\left(  \cosh^2( {\eta / \sqrt{2}})-1-4 e^{6\varphi_1} -
8 e^{3\varphi_1}\cosh({\eta / \sqrt{2}})\right)\ .
\label{onesg}
\ee

\noindent (ii) We consider the same configuration of gauge vectors, but now
we set 
 \begin{align}
 \td \eta_a = (\eta,\eta,\eta,\eta)/2 \,, \quad
 \td \theta_a = (-\theta,\theta,\theta,\theta) \,, \quad 
 \varphi_2=0 \,.
 \end{align}
The Lagrangian takes the form
 \be
  \cL = \sqrt{-g} \Big[ R - 3  (\p \varphi_1)^2 
-    \tfrac12   (\p \eta )^2 - 2  \sinh^2(\tfrac12 {\eta } ) \big(\p\theta +  L^{-1}  A\big)^2  - \tfrac14  e^{4\varphi_1} F_{\mu\nu} F^{\mu\nu}
- V \Big] \,,
\label{lsg2}
\ee
where
\be
V= -\frac{2e^{-4\varphi_1 }}{L^2}\left(   (4 e^{3\varphi_1}-1)  \cosh^2( \tfrac12 {\eta })+1+2 e^{6\varphi_1} \right)\ .
\label{onesg2}
\ee


\subsubsection*{Two equal charges (II)}

We now consider configurations with gauge vectors
 \begin{align}
  A_i = (A,A,0) / \sqrt{2} \,. 
 \end{align}
Similar to the previous case, it is consistent to set the scalars to the values \eqref{scalars} except for $ \td \theta_a = (\theta,0,0,\theta)$.
The Lagrangian takes the form
\begin{align}
\cL = \sqrt{-g} \Big[ R - 3 (\p \varphi_1)^2 
-  \tfrac12  (\p \eta )^2  - \sinh^2(\eta / \sqrt{2}) \big(\p \theta + \sqrt{2} L^{-1} A\big)^2  - \tfrac14  e^{-2\varphi_1} F_{\mu\nu} F^{\mu\nu} 
- V \Big] \,,
\label{ldos}
\end{align}
with $V$ given by (\ref{onesg}).

\subsubsection*{Three equal charges (II)}

Finally we consider the truncation with three equal charges. In this case it is consistent to
set 
 \begin{align} 
\td \eta_a = (0,0,0,\eta) \,, \quad
\td \theta_a = (0,0,0,\theta) \,, \quad
 \varphi_1=\varphi_2=0 \,.
 \end{align}
The remaining scalar $\eta_4$ has the highest charge under this diagonal $U(1)$.
The resulting Lagrangian is given by
\bea
\cL &=& \sqrt{-g} \Big[ R  -  \tfrac12 (\p \eta )^2 - \tfrac{1}{2}\sinh^2(\eta) \big(\p \theta + L^{-1} \sqrt{3} A   \big)^2  - \tfrac14  F_{\mu\nu} F^{\mu\nu} 
- V \Big] \,,
\label{Gubs}
\eea
with
\be
V= \frac{3}{L^2}\cosh^2 (\tfrac12 {\eta })\, \left( \cosh(\eta )-5\right)\ .
\label{VGubs}
\ee
We recognize the Lagrangian obtained in \cite{Gubser:2009qm} by a consistent truncation from IIB theory based on D3-branes at the tip of a Calabi-Yau cone.
We can see explicitly how that model (in the case of $S^5$) also arises from a consistent truncation of ${\cal N}=8$ supergravity. In short, it corresponds to modding out by a $\mathbb{Z}_4$ symmetry and moreover retaining a diagonal $U(1)$ gauge field and a single complex scalar. 
The emergence of the model of   \cite{Gubser:2009qm} in the present context is expected. The reason  is that
the complex scalar field in \cite{Gubser:2009qm} lives within an $\mathcal{N}=1$ supergravity model and is the scalar which is charged under the $U(1)$ of $R$-symmetry; in the present context, the complex scalar field $\eta_4$ is the only one that is charged under the $U(1)$ of $R$-symmetry --the diagonal subgroup of  the three $U(1)$'s that we are retaining in the truncation-- and at the same time neutral under the rest of this gauge group.

\medskip

\begin{table}[ht]
\begin{center}
\begin{tabular}{||c||c|c||c|c||}
\hline 
 & \multicolumn{2}{|c||}{Sector I} & \multicolumn{2}{|c||}{Sector II} \\ \hline
\# Equal charges & $qL$ & $m^2 L^2$  & $qL$ & $m^2 L^2$  \\ \hline
 \hline
One & $2$ &  $-4$   & $1$ & $-3$  \\ \hline
Two & $\sqrt{2}$ & $-4$  & $\sqrt{2}$ & $-3$ \\ \hline
Three & $2 / \sqrt{3}$ & $-4$  & $\sqrt{3}$ & $-3$  \\ \hline
\end{tabular}
\end{center}
\caption{The charges and  masses 
of the charged scalars of sectors I and II for different configurations of black hole charges. }
\label{tab:charges}
\end{table}

This concludes our discussion of the different truncations of both non-supersymmetric sectors.  In table \ref{tab:charges}  we summarize masses and charges for the two different truncations. We have always normalised the phase $\theta$ in such a way the first term in the expansion at small  $\eta $ has coefficient 1/2, i.e.  $\tfrac12 \eta^2 (\p \theta)^2+...$. This leads to a universal normalisation for the term 
$\tfrac12 q^2 \eta^2 A^2$  coming from $(\p\theta +q A)^2$. 
Note that the charge of the scalar field $\eta$ of sector I decreases as one includes more BH charges, while in sector II this is exactly opposite: the charge of $\tilde \eta$ increases with the BH charges. The masses are common in both sectors. 
In the next section we will discuss the interplay between sectors I and II in the context of condensed matter applications.

\section{Thermodynamics and condensed matter applications}
\setcounter{equation}{0}

The truncations obtained in the previous section can be described in terms of   the following general Lagrangian
\be
  \cL = \sqrt{-g} \bigg( R - 3(\p \varphi)^2 -  \tfrac14 G(\varphi) F^{\mu\nu}F_{\mu\nu}
- \tfrac{1}{2}  \p_\mu\eta \p^\mu \eta -  \tfrac{1}{2} J(\eta) A_\mu A^\mu -V(\eta,\varphi) \bigg)\ ,
\label{kkl}
\ee
where the coupling functions $G(\varphi), J(\eta)$ and the potential $V(\eta, \varphi)$ determine the specific truncation. 
In the one-charge and two-equal charge cases, the dilaton $\varphi$ cannot be decoupled and therefore one needs to investigate dilatonic black hole solutions. In the  
 three-equal charge case, the Lagrangian (\ref{isot}) and (\ref{Gubs}) are obtained upon  setting $\varphi=0$ in (\ref{kkl})
and choosing the appropriate $J(\eta)$ and $V(\eta)$. 
The  above Lagrangian provides a useful setup to package the dynamics of the various models. The field $\eta $ in (\ref{kkl}) will denote the  normalized charged field in either sectors I or II. 
We will return to the original notation $\td \eta $  for the sector II charged scalar whenever
this distinction is relevant.

\subsection{Equations of motion  and asymptotic behavior}

Given the comprehensive setup (\ref{kkl}), our ansatz for the metric and the gauge field  is 
\be\label{AnsatzEQM}
ds^2=-g(r)e^{-\chi(r)}dt^2+\frac{dr^2}{g(r)}+r^2(dx^2+dy^2+dz^2),\qquad A=\Phi(r)dt \ ,
\ee
with the scalars fields  being functions only of the radial coordinate.
The equations of motion derived from the Lagrangian (\ref{kkl}) are then given by
\bigskip
\bea
\frac{3}{2}\chi'+3 r \varphi'^2 + \frac{r}{2}\eta'^2+\frac{r}{2 g^2}e^{\chi}J(\eta)\Phi^2=0\ ,\label{Eqchi}\\
\nonumber \pagebreak[4] \\
3\Big(\frac{g'}{gr}+\frac{2}{r^2}\Big)+\frac{1}{2}\Big(\eta'^2+J(\eta)e^{\chi}\frac{\Phi^2}{g^2}\Big)+3\varphi'^2+\frac{e^{\chi}}{2 g}G(\varphi)\Phi'^2+\frac{V(\eta,\varphi)}{g}=0\ ,\label{Eqg} \pagebreak[4] \\
\nonumber\\
\Phi''+\Phi'\Big(\frac{3}{r}+\frac{\chi'}{2}+\frac{\p_{\varphi}G}{G}\varphi'\Big)-\frac{J(\eta)}{gG(\varphi)}\Phi=0\ ,
\label{Eqphi}\\
\nonumber \pagebreak[2] \\
\varphi''+\varphi'\Big(\frac{3}{r}-\frac{\chi'}{2}+\frac{g'}{g}\Big)+\frac{e^{\chi}}{12g}\p_{\varphi}G\ \Phi'^2-\frac{\p_{\varphi}V(\eta,\varphi)}{6g}=0\ ,\label{Eqvphi}\\
\nonumber\\
\eta''+\eta'\Big(\frac{3}{r}-\frac{\chi'}{2}+\frac{g'}{g}\Big)+\frac{e^{\chi}}{2g^2}\p_{\eta}J\ \Phi^2-\frac{\p_{\eta}V(\eta,\varphi)}{g}=0\ .\label{Eqeta}\\
\nonumber
\eea

We are interested in black hole configurations that have a regular event horizon and appropriate boundary conditions at infinity. 
It is easy to check that all  truncations given above possess an AdS vacuum for which $\eta=0$, $\varphi=0$, $\Phi=0$. Therefore, in order to apply the AdS/CFT dictionary we will
look for asymptotically AdS black holes and  require that $g\approx r^2/L^2$ as $r$ goes to infinity. 
Linearizing the equations around this AdS vacuum we find that a generic solution has
the following asymptotic behavior
\bea
\varphi &=& \frac{O_{\varphi}}{r^2}+\frac{C_{\varphi}}{r^2}\log r+\ldots \\
\nonumber\\
\Phi &=& \mu-\frac{\rho}{r^2}+\ldots\label{Asy}\\
\nonumber\\
\chi &=& \chi_{\infty}+\ldots\\
\nonumber\\
e^{-\chi}\, g(r)&=& e^{-\chi_\infty}\, \left({r^2\over L^2}-{\eps\over r^2}+\ldots\right)\ ,
\eea
where the dots stand for terms of higher order in the expansion in powers of $1/r$.
The asymptotic of the dilaton is universal in all models; indeed in all cases $\varphi$ is a scalar field in AdS with $m^2L^2=-4$.
On the other hand, the charged scalars have different masses in sector I and in  sector II. 
In sector I these charged scalars have $m^2L^2=-4$ and they have the same asymptotic behavior  as  the dilaton,
\be
\eta = \frac{O_{\eta}}{r^2}+\frac{C_{\eta}}{r^2}\log r+\ldots \\
\ee
while in the sector II these correspond to scalars with $m^2L^2=-3$, therefore
\be
\tilde \eta = \frac{C_{\eta}}{r}+\frac{O_{\td \eta}}{r^3}+\ldots \\
\ee

The location of the horizon $r=r_h$ is defined by the simple zero of $g$ lying at the largest $r$. The Hawking temperature of the black hole can then be calculated 
as usual by the formula,
\be
T_{\rm Hawk}=\frac{1}{4\pi}g'(r)e^{-\chi(r)/2}\Big|_{r=r_h}\ .
\ee 
The value of $g'(r_h)$ is determined from the first order equation (\ref{Eqg}), in particular by the combination 
\be
g'(r_h)=-r_h\Big[\ V(\eta_h,\varphi_h) + \tfrac12 {e^{\chi_h}}G(\varphi_h)E_h^2\ \Big]\ ,
\ee
where we have defined the parameters
\be
\eta(r_h)=\eta_h, \qquad \varphi(\r_h)=\varphi_h, \qquad \Phi'(r_h)=E_h,\quad \mathrm{\ and}\qquad \chi(r_h)=\chi_h\ .
\ee
Finding an analytic solution of the complete system (\ref{Eqchi})-(\ref{Eqeta}) is, in general, a very difficult task. The particular case 
$\eta \equiv 0$ is much  simpler because the Maxwell equations can then be integrated and the solution can be substituted into the other equations. In particular, for the non-supersymmetric truncations that we are considering, analytic solutions do exist and are described by the ``STU" black hole \cite{Behrndt:1998jd}. Explicit formulas are given in the next section. At any rate, the complete system (\ref{Eqchi})-(\ref{Eqeta}), including the back-reaction of the geometry, can be generically solved using numerical methods, including  the case $\eta \equiv 0$.

The strategy is as follows. We specify the initial data at the horizon in such a way that the Cauchy problem (\ref{Eqchi})-(\ref{Eqeta}) is well posed and then numerically integrate up to infinity. A priori, we have nine parameters to deal with: the location of the horizon $r_h$ and eight initial conditions for the equations of motions. The requirement $g(r_h)=0$ constrains some of them. In particular, the consistency of equations (\ref{Eqvphi}) and (\ref{Eqeta}) fixes the values of $\eta'(r_h)$ and $\varphi'(r_h)$ to satisfy the relations
\bea
g'(r_h)\eta'(r_h)& = & \p_{\eta}V(\eta_h,\varphi_h)\ , \\
\nonumber\\
g'(r_h)\varphi'(r_h)& =&\tfrac16 {\p_{\varphi}V(\eta_h,\varphi_h)}-\tfrac{1}{12} {e^{\chi_h}} \p_{\varphi}G(\varphi_h)\ E_h^2\ ,
\nonumber
\eea
whereas the condition $\Phi(r_h)=0$ is needed to ensure that the gauge field is well defined at the horizon.
Therefore, out of the nine parameters we started with, only the ones in the set $\{ r_h, \eta_h,\varphi_h,E_h,\chi_h\}$ are independent.
This set can be further reduced. In order to do so, 
we note the following two scaling symmetries,
\be
r\rightarrow a r\ ,\qquad (t,\vec{x})\rightarrow a^{-1}(t,\vec{x})\ ,\qquad g\rightarrow a^2 g\ ,\qquad \Phi\rightarrow a\Phi\ ,
\ee
and
\be
e^{\chi}\rightarrow a^2 e^{\chi}\ ,\qquad t\rightarrow a t\ ,\qquad \Phi\rightarrow a^{-1}\Phi\ ,
\ee
which leave invariant the metric, the gauge field and the equations of motion. These two symmetries can be used to set $r_h=1$ and $\chi_{\infty}=0$. Thus, each black hole solution obtained by integrating the equations of motion from the horizon is characterized in terms of the three horizon parameters $\eta_h$, $\varphi_h$, $E_h$.

Finally, the relevant configurations will be the ones with the asymptotic constraints $C_{\varphi}=0$ and $C_{\eta}=0$ (or $C_{\td\eta}=0 $). These two additional conditions leave a one-parameter family of hairy black hole solutions, where the parameter characterizing the solution can be taken to be the temperature. These configurations  represent the physical systems that we are going to describe.

\medskip

It should be noted that in all  truncations it is consistent to look for solutions with  vanishing $\eta$ (which, in particular, implies $C_{\eta}=0$ or $C_{\td\eta}=0 $). Therefore there will be two kinds of interesting configurations: hairy black holes, with a non-trivial $\eta$ turned on, and ``bald" black holes with $\eta\equiv 0$.

\medskip

The field theory interpretation of these two types of solutions follows from the standard AdS/CFT dictionary, that we now briefly review.
This asserts that each field in the gravitational action is dual to a certain operator in the conformal field theory. In five dimensions the correspondence between a scalar field in the bulk with some operator in the CFT of conformal dimension $\Delta $ is given by the relation 
$m^2L^2=\Delta(\Delta-4)$.  
Consider, for concreteness, the scalar $\tilde{\eta}$ of sector II. Its mass is given by $m^2L^2=-3$, therefore there are two solutions: $\Delta_-=1$ and $\Delta_+=3$. This fact shows up in the powers of $r$ arising from the behavior  of the field at infinity, 
\be
\tilde \eta = \frac{C_{\eta}}{r}+\frac{O_{\td \eta}}{r^3}+\ldots\nonumber\\
\ee  
Among the two terms, only the mode associated with $\Delta_+ = 3$ is normalizable; thus, one interprets $O_{\td \eta}$ as the condensate associated with an operator of dimension $3$ in the presence of the external source $C_{\eta}$. Demanding that the solution has $C_{\td\eta}=0$ implies that
the field theory is not sourced by this operator; a non-zero value of $O_{\td \eta}$  then implies a vacuum expectation value
for the CFT operator of dimension 3. In other words the AdS/CFT correspondence implies that a hairy black hole with appropriate boundary conditions is dual to a condensed phase of the system, while a ``bald" black hole represents an uncondensed phase.
Moreover, because $\tilde{\eta}$ is charged under a $U(1)$ symmetry,  its  dual operator will carry the same charge under this symmetry and
a vacuum expectation value of this operator then implies $U(1)$ spontaneous symmetry breaking.

For the charged scalar $\eta $ of sector I there is a unique solution to the mass/dimension relation. This scalar   is dual to a charged operator of dimension $\Delta=2$. The dilaton  is dual to an operator of the same dimension, but neutral under the $U(1)$ symmetry. In these cases, when $m^2=m^2_{BF}$, a logarithimic branch appears in the asymptotic. Such a branch will necessary introduce an instability unless it is treated as a source \cite{Horowitz:2008bn}. This means that one must set $C_\eta =0$ and the interpretation then follows as for the $\tilde{\eta}$ field.

\subsection{The uncondensed phase}

In the maximal $SO(6)$ gauged supergravity, black hole solutions carrying arbitrary charges with respect to three different $U(1)$ symmetries are described by the ``STU" black hole. The present truncations contain these black holes since they maintain the $U(1)\times U(1)\times U(1)$ gauge symmetry and thus the three relevant vector fields. 
Indeed, keeping only the real scalar fields in sector I and sector II, we recover the Lagrangian of \cite{Behrndt:1998jd},
\be\label{STULagrangian}
\cL = \sqrt{-g} \Big( R -   \sum_{i=1}^3 X_i{}^{-2} \Big[\tfrac12 (\p X_i)^2  + \tfrac14  F^{i,\mu\nu} F^i_{\mu\nu} \Big]\Big)\ ,
\qquad\qquad X_1X_2X_3=1\ ,
\ee
for which the following analytic solution was found:
\bea
&&ds^2=-f\, H^{-2/3}\, dt^2+H^{1/3}\, f^{-1}\, dr^2+H^{1/3}\, \frac{r^2}{L^2}\, d\vec{x}^2 \ ,
\nn\\
\label{STUBH}
&&A_i=\left(\frac{Q_i\, \sqrt{m}}{r_h^2+Q_i^2}-\frac{Q_i\, \sqrt{m}}{r^2+Q_i^2}\right) dt\ ,\qquad X_i=H_i^{-1}\, H^{1/3}\ ,
\\
&&
f=\frac{r^2}{L^2}\, H-\frac{m}{r^2}\ ,\qquad H=H_1\,H_2\,H_3\ ,\qquad H_i=1+\frac{Q_i^2}{r^2}\ .\nn
\eea
%
The charged scalar fields are vanishing, therefore these solutions describe the uncondensed phase for our models. The solution (\ref{STUBH}) can be written in the form (\ref{AnsatzEQM}) by a coordinate transformation on the metric: $r^2H^{1/3}\leftrightarrow r^2$.
It is  easy to check that the solution  (\ref{STUBH}) satisfies the equations of motion  (\ref{Eqchi})-(\ref{Eqeta}).
For the case of three equal charges, the coordinate transformation simply becomes $\ r^2+Q^2\leftrightarrow r^2\ $ and one recovers the standard form of the AdS Reissner-N\" ordstrom black hole.

\medskip

It is useful to review the relevant quantities for the field theory thermodynamics in the uncondensed phase (see \cite{Gubser:2009qt,Aprile:2010ge} for  recent discussions).
The energy, entropy and charge densities are
\be\label{TermoQuant}
\hat\epsilon = {3m\over 8\pi^2 L^6}\ ,\qquad \hat s={r_h\sqrt{m}\over 2\pi L^5}\ ,\qquad \hat \rho_i ={Q_i\sqrt{m }\over 4\pi^2L^5}\ ,
\ee
where the quantities $m$ an $Q$, in general, can be read from the asymptotics of the solution.
The position of the horizon is found by solving the equation $f=0$. In general there will be three roots.
We use the notation $r_h^2, \ r_1^2$ and $-r_0^2$ for these roots indicating with $r_h^2$ the greatest (real) one.
The event horizon is then located at $r_h $ and the temperature is given by
\be
T_H={(r_h^2+r_0^2)(r_h^2-r_1^2)\over 2\pi L^2\sqrt{(r_h^2+Q_1^2)(r_h^2+Q_2^2)(r_h^2+Q_3^2)}}\ .
\ee

While the STU black hole represents an exact solution for arbitrary
choice of the $(Q_1, Q_2, Q_3)$,  we will be interested in three particular cases:
one charge, two-equal charges and three-equal charges.
In what follows we show how the uncondensed black holes solutions of sector I and sector II are obtained from the STU black hole solution in each case. 
Finally we observe that having the analytic solution for the uncondensed phases allows us to check the thermodynamical quantities provided by the numerics.

\subsubsection*{Black hole with one single charge}

The first case of interest is when $Q_i = (0,0,Q)$. The solution (\ref{STUBH}) is characterized by the two functions 
\be
H=\left(1+\frac{Q^2}{r^2}\right)\qquad\mathrm{and}\qquad A_3=\left(\frac{Q_i\sqrt{m}}{r_h^2+Q_i^2}-\frac{Q_i\sqrt{m}}{r^2+Q_i^2}\right)dt\ .
\ee
This black hole represents the uncondensed solution for the Lagrangians (\ref{dosnu}) and (\ref{lsg}), in which we set $A_3=A$. The charge density associated to the $U(1)$ gauge group is $\hat\rho=Q\sqrt{m}/(4\pi^2L^5)$.

\subsubsection*{Black hole with two equal charges}

In this case two gauge fields are identified and we set $Q_i=(Q,Q,0)$ in (\ref{STUBH}). The black hole is then specified by the functions
\be
H=\left(1+\frac{Q^2}{r^2}\right)^2\qquad\mathrm{and}\qquad A_1=A_2\ .
\ee
This is a solution to the equations of motion of the Lagrangians (\ref{dosdi}) and (\ref{ldos}) with the identifications
\be
A_1=A_2=A/\sqrt{2},\qquad\qquad X_1=X_2=e^{\varphi_1}\ .
\ee
The charge density associated to the diagonal $U(1)$ is therefore $\hat\rho=Q\sqrt{2m}/(4\pi^2L^5)$.

\subsubsection*{Black hole with three equal charges}

This is the case when $Q_i=(Q,Q,Q)$ and the three gauge fields are identified. Now we have
\be
H=\left(1+\frac{Q^2}{r^2}\right)^3\qquad\mathrm{and}\qquad A_1=A_2=A_3\ .
\ee 
This represents a solution for the  uncondensed phase  for the models (\ref{isot}) and (\ref{Gubs}) with the identifications
\be
A_1=A_2=A_3=A/\sqrt{3}\ ,\qquad\qquad X_1=X_2=X_3=1\ .
\ee
The charge density then is given by $\hat\rho=Q\sqrt{3m}/(4\pi^2L^5)$.

\medskip

Let us now discuss some basic aspects of the thermodynamics at fixed $\hat\rho$. 
 The Hawking temperature is given by the formula
\be\label{OtherT}
T=\frac{1}{4\pi}H(r_h)^{-1/2}f'(r_h)\ .
\ee 
{}The horizon equation, $f=0$, gives
\be\label{Otherhorizon}
m=\frac{r_h^4}{L^2} H(r_h)\qquad \mathrm{where}\qquad H=\left(1+\frac{Q^2}{r_h^2}\right)^{\alpha}\qquad\alpha=1,2,3\ ,
\ee
with $\alpha$ counting the number of equal charges. It is convenient way to rescale  $r\to r_h \tilde r$, so the horizon is at $\tilde r=1$.
We also introduce new parameters $\td m$ and $\td Q$ by the rescaling
\be
m= r_h^4 \tilde m\ ,\qquad\qquad Q= r_h\tilde Q\ .
\ee 
Then the relation (\ref{Otherhorizon}) simplifies, reducing to the formula $\tilde m=\tilde H/L^2=(1+\tilde Q^2)/L^2$. 
Using these relations, the temperature (\ref{OtherT}) becomes
\be
T=\frac{r_h}{2\pi L^2}(1+\tilde Q^2)^{1-\alpha/2}\left(2+(2-\alpha)\tilde{Q}^2\right)\ .
\ee
We are interested in  the thermodynamics in the fixed $\hat\rho$ thermal ensemble. 
In terms of the rescaled variables, the charge density $\hat\rho$ takes the form
\be
\hat\rho=\frac{\tilde Q\sqrt{\alpha\ \tilde m}}{4\pi^2 L^5}r_h^3\ =\ \sqrt{\alpha}\ \frac{\tilde Q\ (1+\tilde Q^2)^{\alpha/2}}{4\pi^2 L^6}r_h^3
\ee
Thus, solving for $r_h$, the temperature becomes
\be
T=\frac{\tilde{Q}^{-1/3}}{(2\pi\sqrt{\alpha} )^{\frac{1}{3}}}\left(2+(2-\alpha)\tilde{Q}^2\right)\left(1+\tilde Q^2\right)^{\frac{\alpha}{3}-1}
\hat\rho^{\frac{1}{3}}\ .
\ee 
Since $\hat \rho$ is fixed, the temperature $T$ is only a function of the auxiliary variable $\tilde Q$. {}From this expression
one can get a qualitative understanding of the properties enjoyed by the different ensembles. 
In the one charge case $\alpha=1$ and $T$ is a strictly positive function that goes to infinity as $\tilde Q$ approaches the two limits, $\tilde Q\ll 1$ and $\tilde Q\gg 1$, therefore the uncondensed phase has a minimum temperature whose significance was extensively discussed in \cite{Aprile:2010ge}.
We shall see that the system can be driven to lower temperatures by condensation; in other words, there are black hole solutions with hair that  reach lower temperatures. 
In the two-equal charge ensemble one has $\alpha=2$ and $T$ varies between $0$ and infinity. 
Some  aspects  of the thermodynamics of these solutions are discussed in \cite{Gubser:2009qt}. For the purpose of this work, we
point out that this black hole reproduces some features of a Fermi-liquid phase and therefore the search for the condensed phase is well motivated. 
Some of these features are, in particular, the linear specific heat and zero entropy as $T\rightarrow 0$. 
In the AdS Reissner-N\" ordstrom black hole, $\alpha=3$ and $\tilde Q$ is restricted to be in the interval  $\tilde{Q}\in [0,\, \sqrt{2}]$. The temperature vanishes as $\tilde Q\to \sqrt{2}$, but the horizon has a finite size resulting in a large value for the entropy at zero temperature.

\subsection{The condensed phase}

In what follows we will look for a condensed phase in the different sectors. These truncated models could be regarded as holographic setups for AdS/CMT applications on their own right. Nevertheless, the most interesting aspect of top-down constructions, like the present one, is represented by the explicit knowledge of the dual field theory\footnote{
 Another approach where the dual field theory is known is based on using D-brane probes in  string-theory black brane backgrounds (see e.g. \cite{Erdmenger:2011hp} and references therein).}.
We hope that this explicit connection will open the way to the study of novel features of the thermodynamics 
of large $N$ SYM in the strong coupling regime. 
On the field theory side, some aspects of its phase diagram have been discussed on $S^3$ 
(see e.g. \cite{Yamada:2006rx,Hollowood:2008gp}) but so far  there have not been many  discussions on the spontaneous symmetry breaking of $U(1)$ symmetries.

In principle, in order to understand the thermodynamics of the system at fixed charge densities, one should search for the dominant thermodynamic configuration not only in a given sector but in  the full ten-dimensional theory.
This would of course be a complicated task already in the context of $D=5$ maximal $SO(6)$ supergravity, because one should look
for the hairy black hole configuration with least free energy among  configurations where any of the 42 scalars can be turned on (or, more generally, even one-form or two-form hair as in \cite{Aprile:2010ge}). Despite this huge number of possibilities, an important glimmer of information is provided by the observation that not all sectors are relevant in order to identify
the highest critical temperature at which the first phase transition can occur.
Indeed, one can show (see e.g. figure 2 in \cite{Gubser:2009qm}) that the critical temperature increases with the charge and decreases with the mass
of a given mode. This reflects the fact that, for given charge, the dual operators with the minimal dimension are those which should
condense first; whereas, for given conformal dimension, operators with highest R-charge should condense first. The competition between charge and dimension will appear in several examples below. Furthermore,  in section 4 we will also  discuss the possibility of condensation arising from some of the 42 scalars not included in our truncated Lagrangians.

The different sectors studied in this paper also include modes\footnote{Some special features in the conductivity arise for scalars saturating the BF bound, see  \cite{Horowitz:2008bn}.}  with lowest possible mass (saturating the BF bound) which are therefore dual to (protected) operators of lowest dimensions. Nonetheless, we shall see that their dual operators  do not always condense before  other operators of the theory: in some cases there are  modes which, while having higher mass, they have  a  sufficiently larger charge to trigger condensation
at a higher critical temperature.

In what follows we shall numerically exhibit  the competition between  sectors I and II. All the plots shown in this section are obtained in the canonical ensemble at fixed charge density $\hat\rho = 1$. For the uncondensed black holes solutions we refer to the formulas already given case by case in section $3.2$. We recall that $\hat\rho=\rho/(4\pi^2 L^5)$, where $\rho$ is the asymptotic value that we read from the expansion (\ref{Asy}) as usual. Similarly for the energy $\hat\epsilon$. In the gauge (\ref{AnsatzEQM}) the entropy and then the free energy density are given by
\be
\hat s={r_h^3\over 2\pi L^6}\ ,\qquad\qquad \hat{f}=\hat{\epsilon}-T\hat{s}\ .
\ee

\subsubsection*{Condensation in single-charge thermal ensemble }

\begin{itemize}

\item {\it Condensation in Sector I}

Consider the Lagrangian (\ref{dosnu}), (\ref{tresnu}).
It describes a complex scalar of charge $qL =2$ and $m^2 L^2 = -4$ and a real scalar $\varphi $ of mass $m^2 L^2 =-4$.
We have numerically computed the critical curve representing the order parameter as a function of the temperature, including backreaction
(see Fig. \ref{fig1charge}a).
We find that there is a critical temperature $T_c\cong 1.04$ below which a hairy black hole appears.
This curve presents some unusual features. 

At some lower temperature $T_1\cong 1.02$ the second derivative of the condensate with respect to the temperature changes sign. This could be an indication of a new phase transition as it
implies that the fourth order term of the free energy in terms of the condensate must have a
strong temperature dependence near $T_1$. In Landau-Ginzburg theory, it is normally assumed that the fourth order coefficient
is not strongly temperature dependent around the critical point $T_c$, but
it is in principle possible that there is some new temperature
scale below $T_c$ where this coefficient also starts to change. In particular, if it goes to zero or 
becomes negative then this term no longer stabilizes the condensate, and higher order coefficients
(if present) become important. Because of this, the behavior of the
condensate can change and indeed can blow up at this new temperature scale, as indeed occurs in the present model.

This  behavior can be seen with more clarity by considering a scalar field with the same Lagrangian but larger charge $q$.
In Fig. \ref{fig1charge}b  we have plotted the condensate as a function of the temperature for $q=16$.
One can see that, near $T_c$, the behavior  is very much like in second order phase transitions of ordinary mean field theories.
Then, at some lower temperature  $T_1^{(q=16)}\cong 0.75$ the second derivative of the order parameter changes sign and
the order parameter then increases very rapidly.
{}For $q\to \infty $ , one reaches the probe limit where the dilaton and metric are decoupled, and one can study the system
in terms of a charged scalar in the AdS Schwarzchild black hole background. In this  limit, $T_1^{(q)}\to 0$, i.e. there is no change of sign in the second derivative of the condensate curve, which has the usual mean field shape all the way down to $T=0$.

Returning to the $q=2$ case, on the gravity side, as the temperature is slightly lowered below $T_1$, the black hole gets
drastically reduced  to a very small size, i.e. $r_h$  becomes
very small, and the internal energy $\hat\epsilon $ of the system
goes down abruptly to almost zero value. This is shown in Fig. \ref{energies}a. 
As a result,  the specific heat has a very large peak near $T_1$.
In the $q=16$ case, the internal energy  also goes to almost zero value, but at a slower rate (see Fig.~\ref{energies}b).
One common feature which is present in both cases is that the condensate is catapulted to large values at temperatures just below  the inflexion point in the critical curves of Fig.~\ref{fig1charge}a,b. Another common feature is the emergence of a hairy black hole with very small entropy and energy at finite $T$. 
Because these quantities are obtained from the asymptotic behavior of the solution, when they get small,  it becomes difficult to
determine them with sufficient accuracy. The numerical results seem to indicate that energy and entropy  remain small and smoothly decrease as the temperature is lowered from $T_1$ to small values.

We have computed the free energy of both the $q=2$ and the $q=16$ hairy black holes and found that these are indeed lower than the free energy of the dilatonic bald black hole describing the uncondensed phase. Therefore it dominates the thermodynamics at $T<T_c$. 
Figure \ref{free1charge}  shows the free energy for the $q=16$ model, which is easier to visualize. One can see that there is an accumulation of points approaching zero free energy, at temperatures just below $T_1^{(q=16)}\cong 0.75$. 
The deep physical reason for this is mysterious to us, but 
it is a consequence of the fact that the internal energy $\hat \eps$ and entropy $\hat s$ get very small near
that point.

It is worth noting that the hairy black hole solutions reach temperatures which are lower than the minimum temperature of the dilatonic black hole describing the uncondensed phase.
For the case $q=2$, the numerical results show that the hairy black hole solution exist up to a new minimum temperature around $0.22$, where it joins an unstable branch coming from higher temperatures.
Nevertheless, already at temperatures lower than $T_1\cong 1.02$ 
it is far from clear that the solution is describing  reliable physics.
It could also be that there is another
condensate at the temperature $T_1$, i.e. there is some other state
with lower free energy that arises exactly at this new temperature.

\begin{figure}[tbh]
\centering
\subfigure{\includegraphics[width=7.5cm]{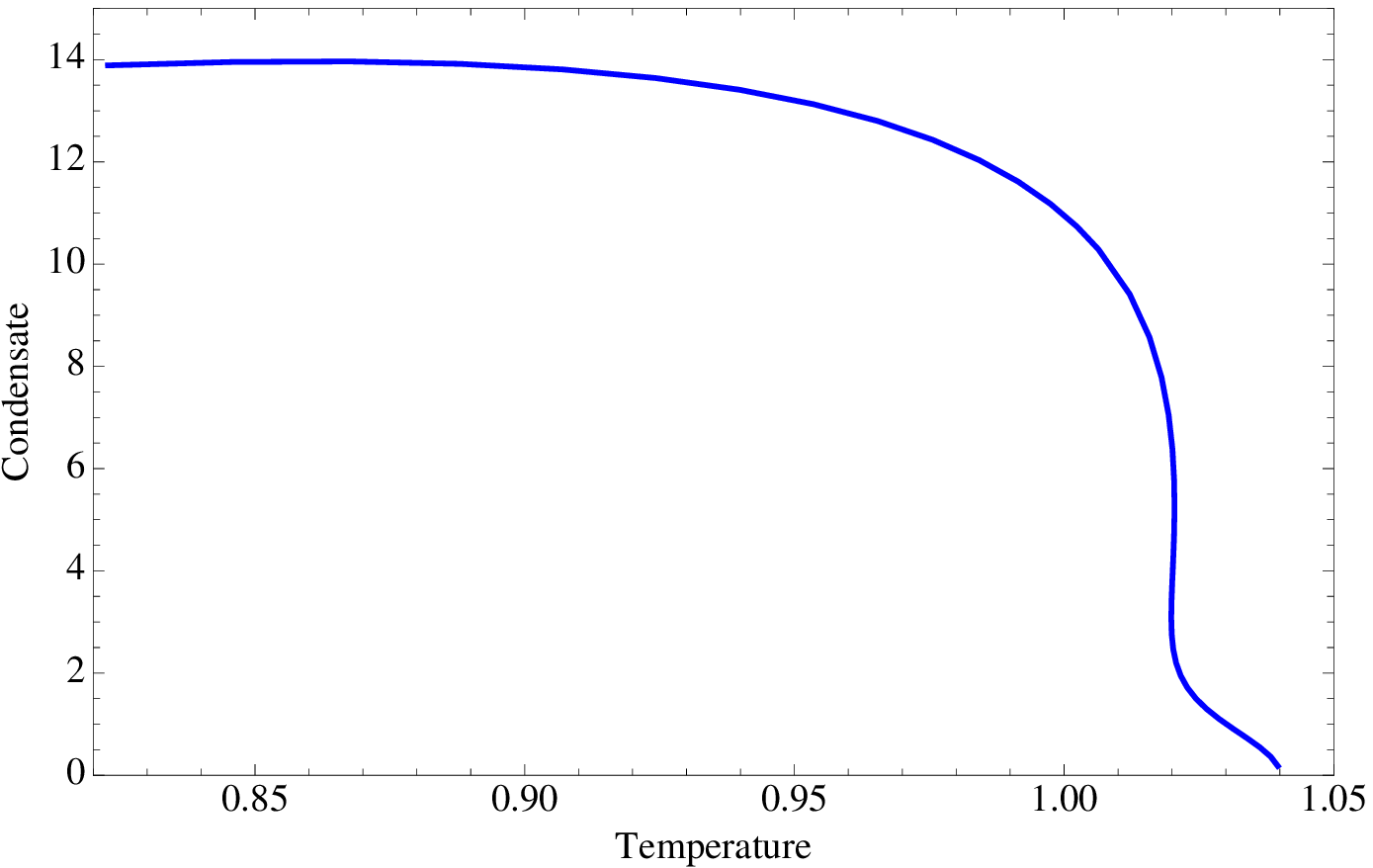}}
\ \ \ \ 
\subfigure{\includegraphics[width=7.5cm]{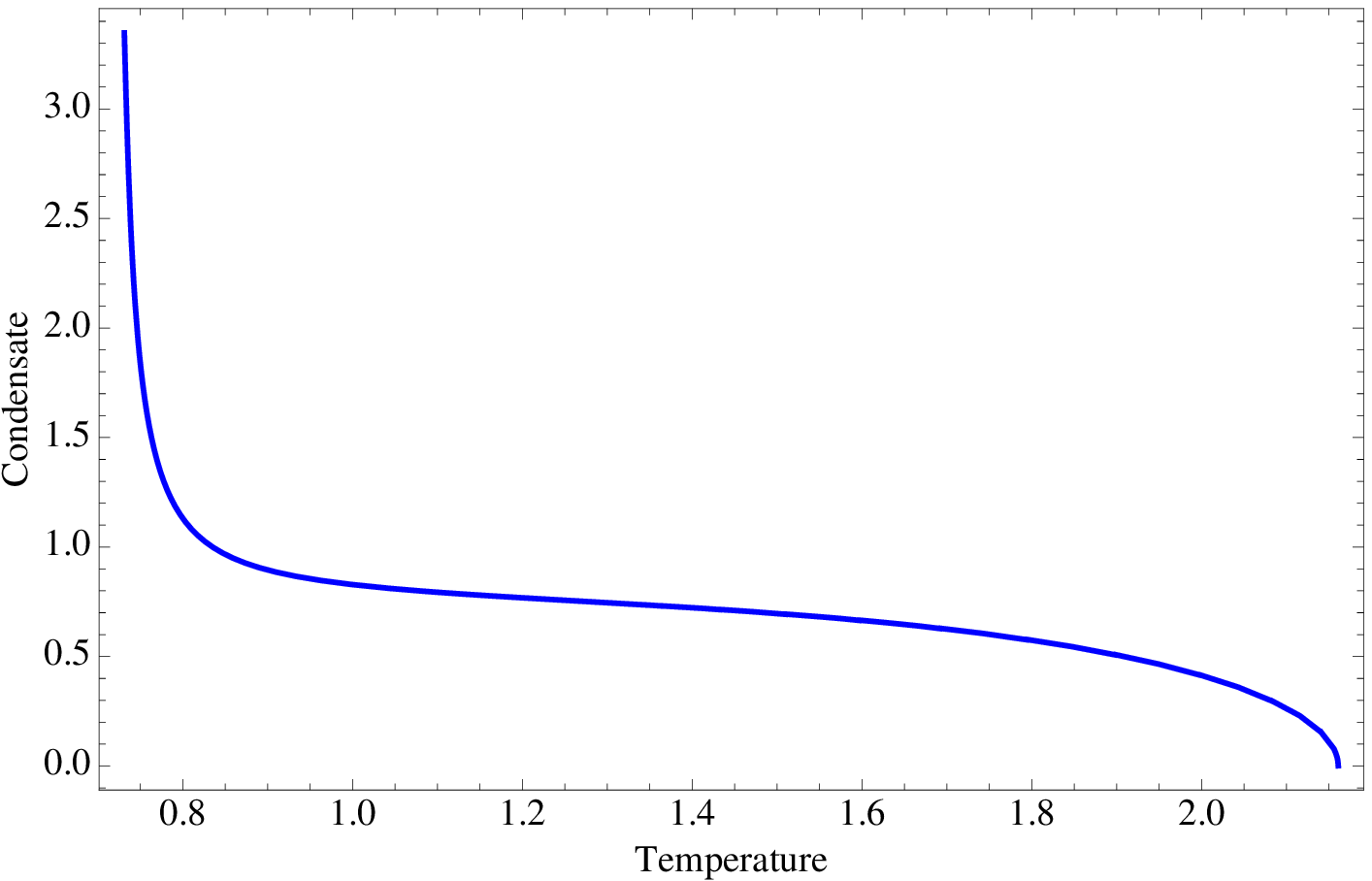}}
\caption{(a) Condensate as a function of the temperature for the one charge model in sector I ($q=2$). 
(b)  Similar plot, for a scalar with the same Lagrangian but charge $q=16$ instead of $q=2$.
\label{fig1charge}}
\end{figure}

\begin{figure}[tbh]
\centering
\subfigure{\includegraphics[width=7.5cm]{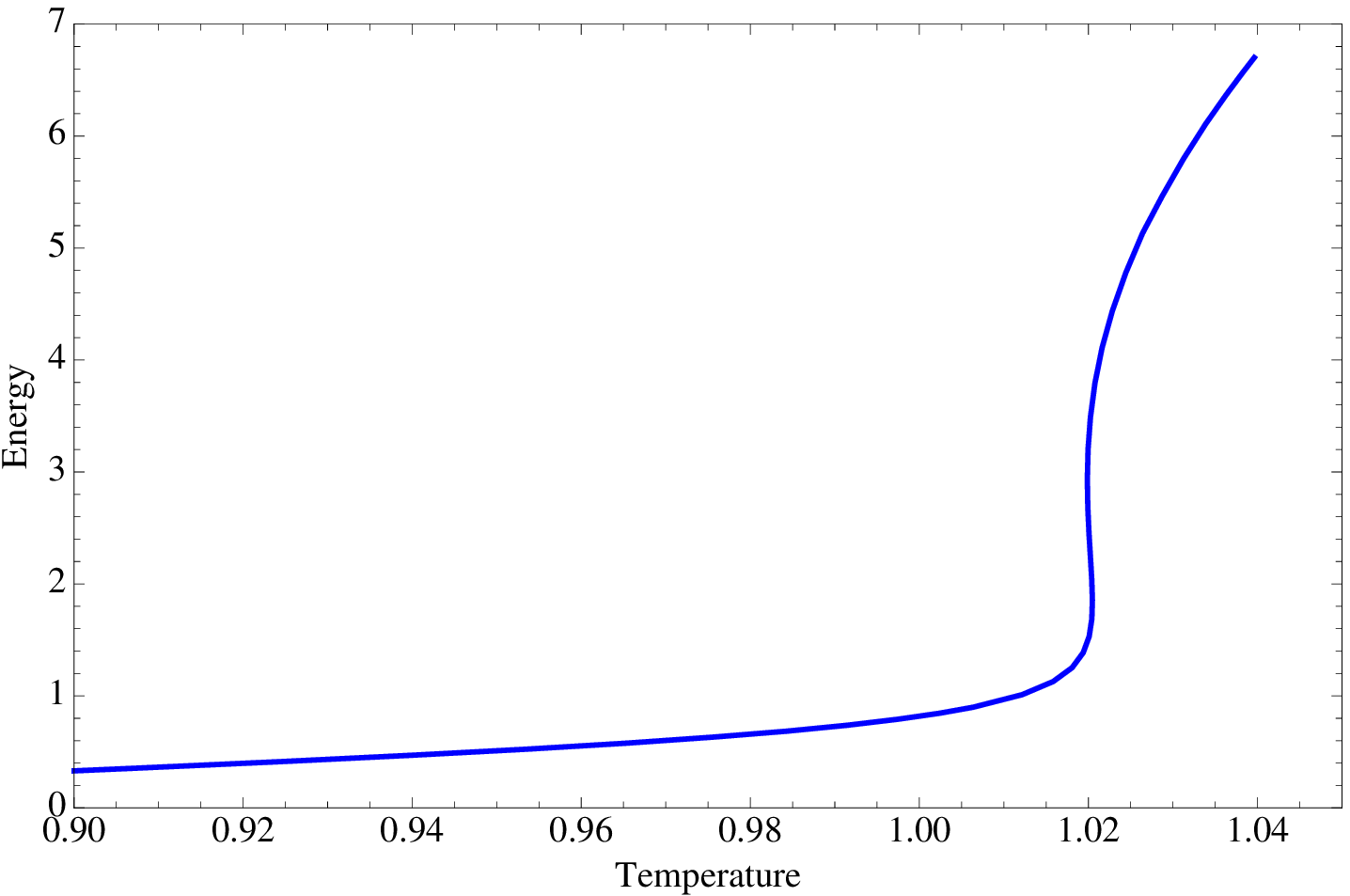}}
\ \ \ \ 
\subfigure{\includegraphics[width=7.5cm]{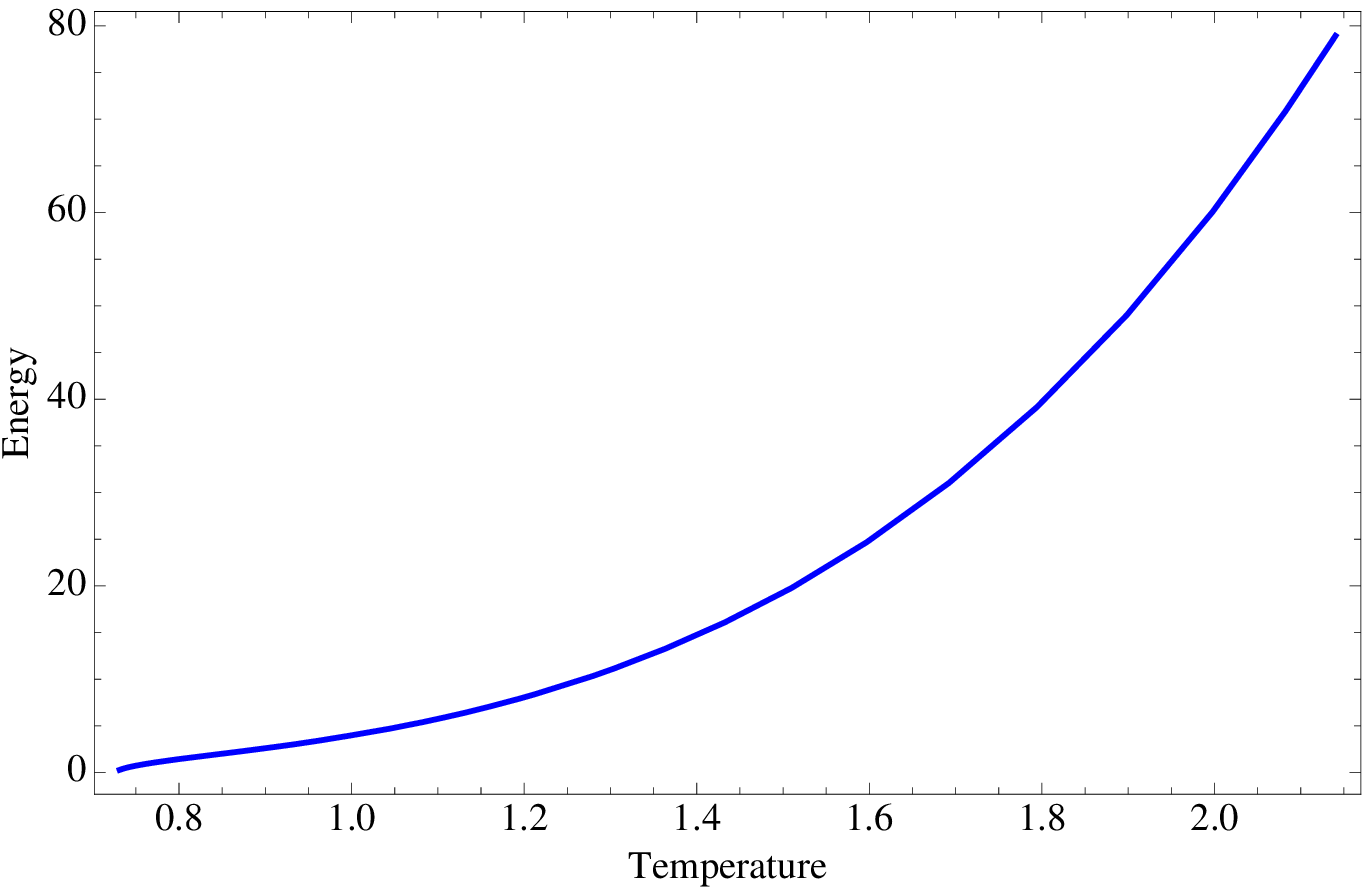}}
\caption{(a) Energy  as a function of the temperature for the one-charge model in sector I.
(b)  Similar plot, for a scalar with the same Lagrangian but charge $q=16$ instead of $q=2$.
\label{energies}}
\end{figure}

\begin{figure}[h!]
\centering
\includegraphics[scale=.7]{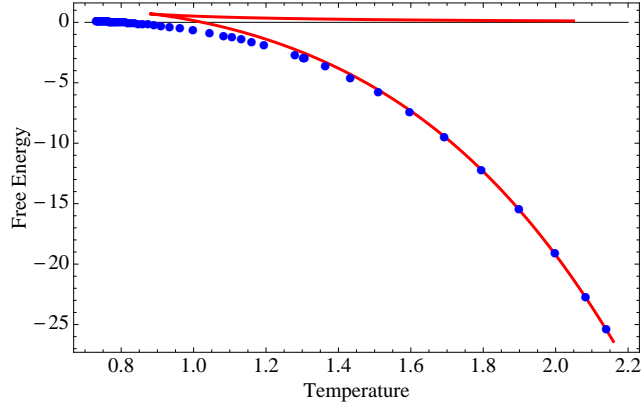} 
\caption{Free energy for the system described by the Lagrangian (\ref{dosnu}), (\ref{tresnu}) with $q=16$ (one charge ensemble, sector I).
 The solid lines represent the two branches of the bald dilatonic black hole solution. The dotted line represents the free energy
 for the hairy black hole solution. 
}
\label{free1charge}
\end{figure}

\newpage

\item {\it Condensation in Sector II}

\noindent Model (i): The relevant Lagrangian is (\ref{lsg}), (\ref{onesg}). It contains a complex scalar field of charge 
 $qL = 1$ and $m^2 L^2 = -3$, plus a real scalar $\varphi$.
Note that it has a lower  charge and larger mass than the previous sector I scalar.
Thus one can expect that, if it condenses, it will be at a lower critical temperature.

By solving the equations numerically we find that for that particular value of the charge there is no condensation, i.e. there is no value for the temperature
at which solutions exist with the required boundary conditions. The origin of this lack of condensation is not only the fact that
the scalar has lower charge and larger mass, but it can also be traced back to
 the coupling between $\eta$ and the dilaton $\varphi$ through the potential $V(\eta,\varphi)$. 
 In the vicinity of the critical temperature, $\eta $ is small everywhere and one can  study the emergence of the hairy black hole by
  the linearized equation for $\eta$ in the bald (dilatonic) black hole background. It has the schematic form: 
\be
\eta''+F(r)\eta'+q^2\eta\ \Phi^2_{\rm uncond.}-m^2\eta\ \tilde V(\varphi_{\rm uncond.})=0\ .
\ee
At infinity the dilaton goes to zero and we have fixed the normalization of $\tilde V$ to be $\tilde V(0)=1$. 
In non-dilatonic models, the critical temperature depends only on $q L$ and $m^2 L^2$. However, in dilatonic models we find that the critical temperature depends also on the specific form of
$\tilde V(\varphi_{\rm uncond.})$. In particular, one can find examples of $\tilde V(\varphi_{\rm uncond.})$ which  give rise to condensation
for a scalar $\eta $ with the same $m^2$ and $q$ as in this model.

\medskip

\noindent Model (ii): Now the relevant Lagrangian is (\ref{lsg2}), (\ref{onesg2}).  As in the previous case, we have a complex scalar field of charge 
 $qL=1$ and $m^2 L^2=-3$, plus a real scalar $\varphi $. We again  find that there is no condensation in this system for similar reasons as in the model (i).

\end{itemize}

\subsubsection*{Condensation in two equal-charge thermal ensemble}

\begin{itemize}

\item {\it Condensation in Sector I}

The relevant Lagrangian is given in (\ref{dosdi}), (\ref{tresdi}).
It contains a complex scalar of charge $qL = \sqrt{2}$ and $m^2 L^2=-4$ and a real scalar $\varphi $. 
Here we find that there exists a family of hairy black holes parametrized by the temperature for any\footnote{A similar
phenomenon was found in four-dimensional models for certain black holes with $AdS_4$ asymptotic in \cite{Buchel:2009ge,Buchel:2011ra}.} $T>T_c$.
On the face of it, this appears to be surprising, since one expects that the condensed phase appears at low temperatures, not at high temperatures.
But for condensation to actually take place, it is necessary that the free energy of this hairy black hole configuration be less
than the free energy of the two-equal charge STU black hole describing the uncondensed phase. 
 We find that the free energy of such hairy black holes is at all temperatures $T>T_c$ 
greater than the free energy of this STU black hole.
Therefore these hairy black holes represent unstable branches that do not contribute to the thermodynamics.
We will refer to this phenomenon of a thermodynamically subdominant condensate at $T>T_c$ as {\it retrograde condensation}\footnote{The term ``retrograde condensation"
was first used by Kuenen in 1892 \cite{Kuenen} to describe the behavior of a binary mixture during isothermal compression above the critical temperature of the mixture (a  discussion can be found in \cite{katz}). Such a system also displays the phenomenon of a subdominant condensate in some temperature range.}.

The results are shown in Fig. \ref{fig2charge}a,b.
One may also wonder if the critical curve could turn back at even higher temperatures not displayed in Fig. \ref{fig2charge}a.
We do not expect that this will happen, because  Fig.  \ref{fig2charge}a includes  temperatures that are high enough to be   above any 
dimensionful scale of the problem and the curve seems to have already reached a well defined asymptotic behavior.

\begin{figure}[tbh]
\centering
\subfigure{\includegraphics[width=7.5cm]{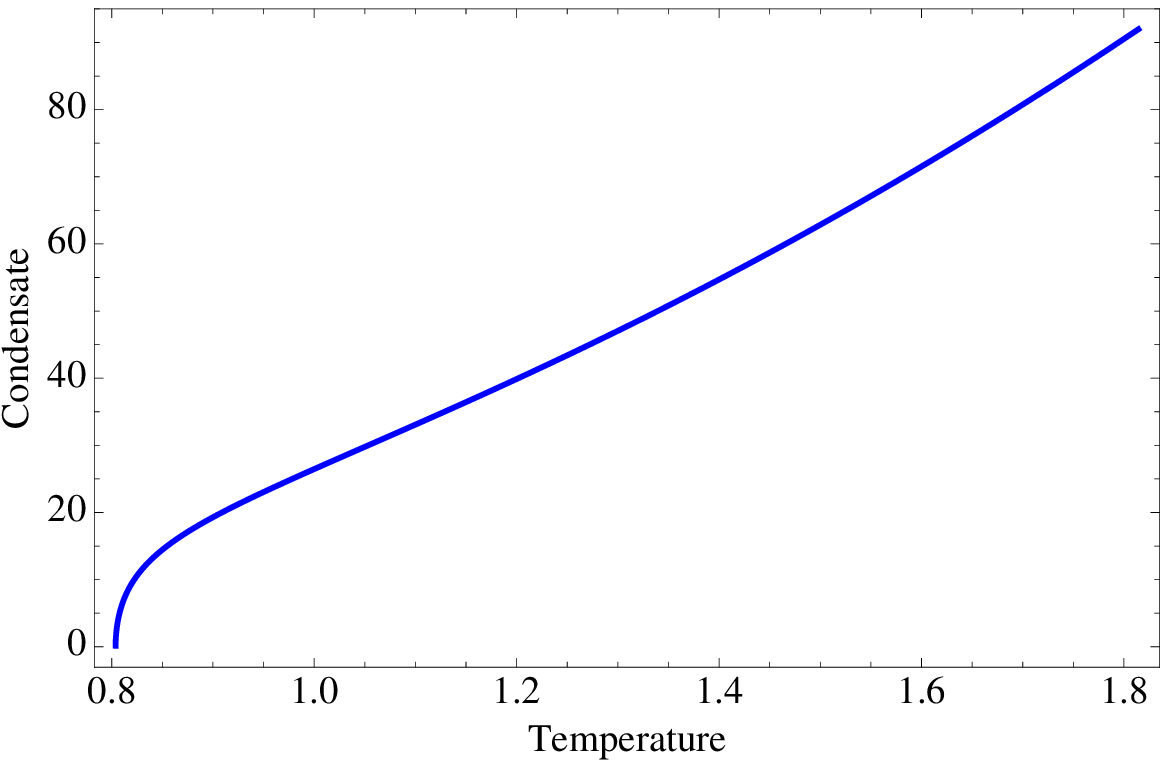}}
\subfigure{\includegraphics[width=7.5cm]{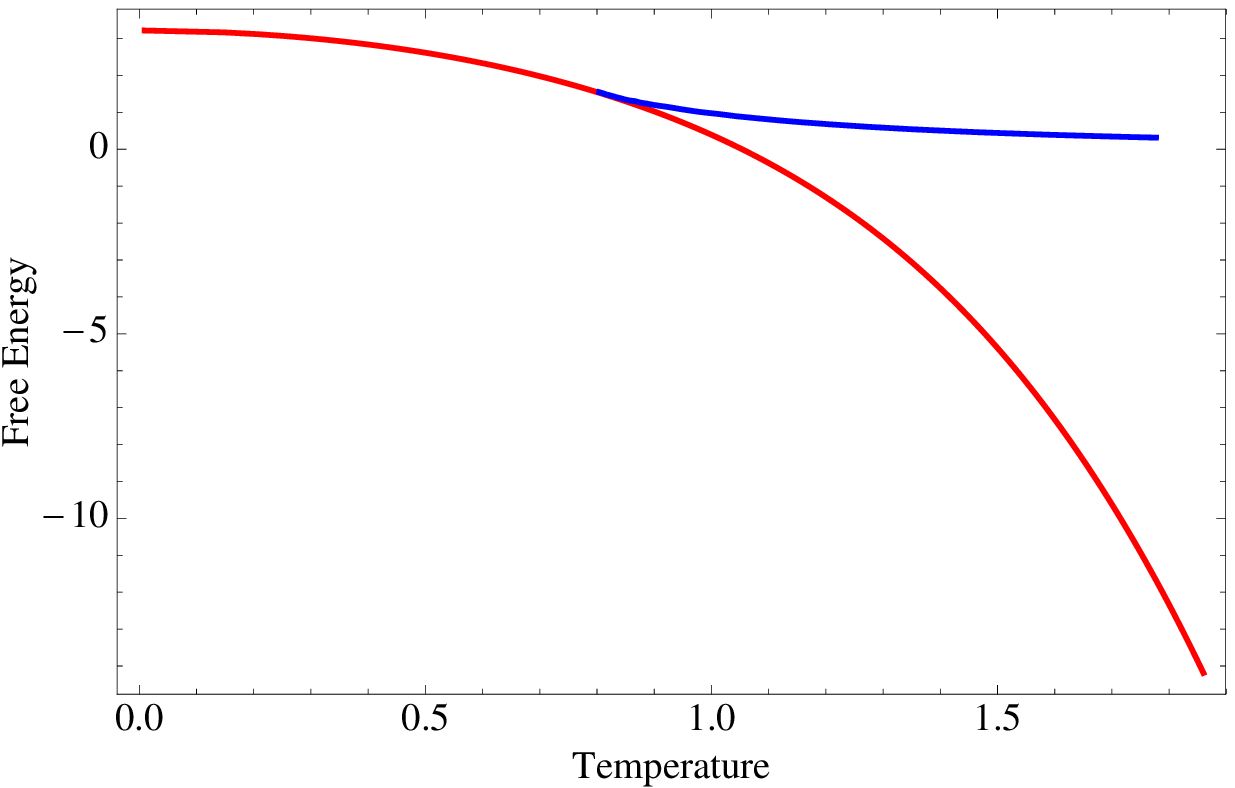}}
\caption{(a) Condensate as a function of the temperature in the two-equal charge thermal ensemble, sector I model.
(b)  Free energy for the same model. The upper (blue) branch is the free energy of the hairy black hole solution, while the lower red branch
is the one for the dilatonic STU black hole with $Q_1=Q_2$, $Q_3=0$.
\label{fig2charge}}
\end{figure}

\item {\it Condensation in Sector II}

We now consider the Lagrangian (\ref{ldos}), with the potential given in (\ref{onesg}).
The complex scalar now has $qL = \sqrt{2}$ and $m^2 L^2=-3$.
Note that  it has lower charge and higher mass than the sector I scalar.
We find that there is no condensation in this model.

\end{itemize}
\subsubsection*{Condensation in  three equal-charge thermal ensemble}

In this case the uncondensed phase is described by the Reissner-Nordstr\" om black hole. The real scalars
$\varphi_1,\ \varphi_2$ are set to zero, which greatly simplifies the analysis. The Lagrangians (\ref{isot}), (\ref{Gubs}) are examples of the general models introduced in \cite{Aprile:2009ai,Franco:2009yz},
\be
  \cL = \sqrt{-g} \big[ R -\tfrac14 G(\eta )\ F^{\mu\nu}F_{\mu\nu}-
\tfrac12  \p_\mu\eta \p^\mu \eta -\tfrac12   J(\eta )  A_\mu A^\mu +{12\over L^2} U(\eta)  \big] \,,
\label{GenMod}
\ee
with the identification
\be
\label{rrr}
G(\eta ) =1\ ,\qquad U(\eta )= \tfrac12   \big(1+ \cosh^2({\eta / \sqrt{3}}) \big)\ ,\qquad J(\eta ) =\frac{4}{L^2}\,   \sinh^2(\eta / \sqrt{3}) \,,
\ee
for sector I, and
 \be
G(\eta ) =1\ ,\qquad U(\eta )= \tfrac14   \cosh^2(\tfrac12 \eta) \big( 5-  \cosh({\eta}) \big) \ ,\qquad J(\eta ) =\frac{3}{L^2}\,   \sinh^2(\eta) \,,
\ee
for sector II.

Some relevant features of the condensation can be learned from the properties of the functions $U(\eta)$ and $J(\eta )$.
We borrow part of the discussion in \cite{Aprile:2010yb,Herzog:2010vz}, originally given for  3+1 dimensional holographic models in the no-backreaction approximation.
Near the critical temperature, $\eta $ is small and one has the expansions
 \bea
  J &=& q^2 \eta^2 + j_0 \eta^4 + \ldots \, ,
 \\
  V &=& - \frac{12}{L^2}U = - \frac{12}{L^2} \big( 1 - \frac{m^2 L^2}{24} \eta^2 - v_0 \eta^4 + \ldots \big) \,. \label{v4}
 \eea
The  quartic terms in $J$ and $V$ play an important role.
This can be understood by using the following formula, deduced in \cite{Aprile:2010yb} (see eq. (3.14)), giving 
the temperature dependence of the order parameter in the vicinity of a second-order transition:
\be\label{MasterEq}
1- \frac{T}{T_c^{(N)}(m^2 ,q^2)} = \langle O_{1}\rangle ^2 (A_N(m^2 ,q^2) + v_0 C_N +j_0 L^2 D_N )+ ... \ .
\ee
Here dots represent terms with higher powers of the order parameter and $A_N, C_N, D_N>0$ are numerical coefficients.
This formula was derived using a series expansion near the horizon, and here $N$ represents the truncation order.
When $v_0=j_0=0$, there are no quartic corrections in $U$ and $J$ functions and the formula reproduces the mean field properties of the condensate as a function of the temperature 
of the HHH model\footnote{
Note that the simplest phenomenological model \cite{Hartnoll:2008kx} with $V=m^2 \eta^2/2$ with $m^2<0$  has
an unbounded potential. Nevertheless, due to the AdS boundary conditions, the model exhibits a  second-order phase transition with standard (mean field type) critical curve
and in particular does not present any runaway behavior.} \cite{Hartnoll:2008kx}. The curve will be qualitatively the same near $T_c$ as long as the coefficient of $\langle O_{1}\rangle ^2 $
is positive. In particular, this is the case if $v_0,j_0>0$.
A drastic change occurs when $v_0 C_N + j_0 L^2 D_N< -A_N $. As we shall see, this condition can be met in models where $v_0$ is sufficiently negative.
Then the condensed phase appears at $T>T_c$, instead of $T<T_c$, at least near $T_c$.
What happens next depends on higher order terms in $V$ and $J$.
In particular, in some cases,  the critical curve  comes back to the lower temperature region $T<T_c$ (examples of this 
behavior, representing first order phase transitions, are in \cite{Franco:2009yz,Aprile:2009ai}).
This depends on higher order terms in the potential so it requires solving the full system of equations including backreaction,
as we do in this paper. In other examples --like in the two-equal charge ensemble discussed above or in the three-equal charge, sector I, discussed below-- the hairy black hole  branch  extends all the way to the region $T>T_c$, and thus displays the phenomenon of retrograde condensation.

\begin{itemize}

\item {\it Condensation in Sector I}

The relevant Lagrangian is given in (\ref{isot}).\footnote{A study of hairy black hole solutions which are asymptotic to {\it global} $AdS_5$
in this sector was carried out in \cite{Bhattacharyya:2010yg}.}
It should be noted that, despite the fact that the potential ($V=-12U/L^2$ with $U$ given in (\ref{rrr})) is unbounded, the theory around the  trivial stationary point is stable due to the fact that in the original ${\cal N}=8$ supergravity  theory  this point is  supersymmetric (which in particular ensures that all scalar fluctuations have masses at or above the BF bound).

We have numerically integrated the equations and found  a family of black hole solutions with charged scalar hair with the correct asymptotic.
Figure \ref{fig3charge}a shows $\langle O_1\rangle$ vs. $T$. Like in the  two equal-charge case for sector I, 
we again find retrograde condensation: the hairy black hole solution exists for $T>T_c$, instead of $T<T_c$ (with $T_c\cong 0.56$).
This is consistent with the fact that the quartic coefficient $v_0$ in the potential is negative.

 This hairy black hole solution again has free energy
which is larger than the free energy of the Reissner-Nordstr\" om black hole. This is shown in
Figure \ref{gubs}. Therefore this condensed phase is unstable and not physically relevant.
 As in the two-charge case, one can also see that the critical curve already reaches some well-defined asymptotic behavior so it is not expected to
turn around.

In the context of Landau-Ginzburg theory, as discussed above, this departure from mean field theory is 
a sign of an unusual temperature dependence of 
the coefficients of higher order terms. For example, 
this behavior might arise if for $T>T_c$ the Landau-Ginzburg potential has a relative maximum at some positive value, which joins the absolute minimum at zero for $T=T_c$. This can be described by a potential with a fourth order coefficient that 
becomes negative for $T>T_c$ (assuming that there are higher order terms that stabilize the potential).

\medskip

In conclusion, there  is no  phase transition in this  sector I.

\begin{figure}[tbh]
\centering
\subfigure{\includegraphics[width=7.5cm]{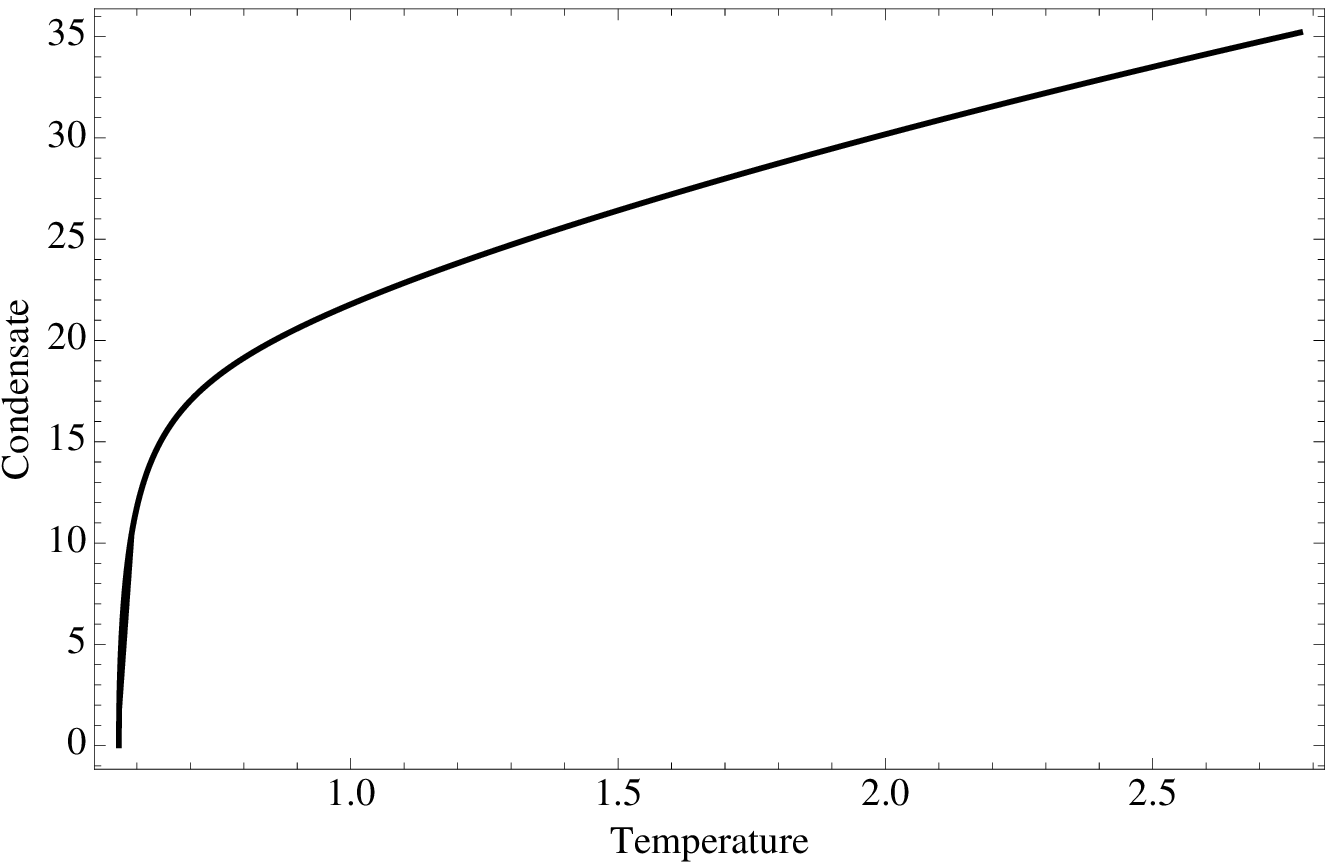}}
\subfigure{\includegraphics[width=7.5cm]{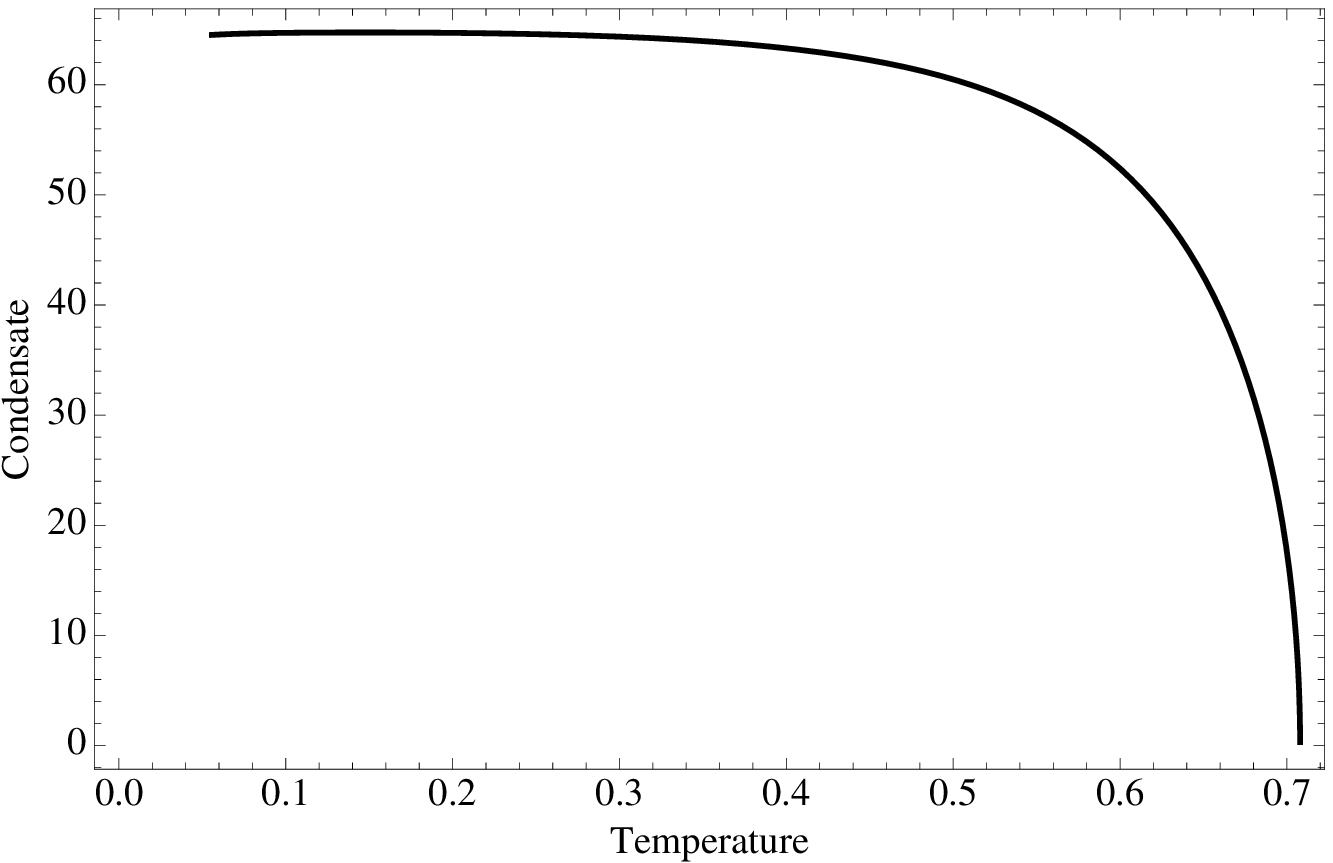}}
\caption{(a) Condensate as a function of the temperature in the three-equal charge thermal ensemble, sector I model.
(b)  
Condensate as a function of the temperature in sector II
(representing the same model of \cite{Gubser:2009qm}, now considered in the thermal ensemble  at fixed charge density rather than at fixed chemical potential).
\label{fig3charge}}
\end{figure}

\begin{figure}[h!]
\centering
\includegraphics[scale=.6]{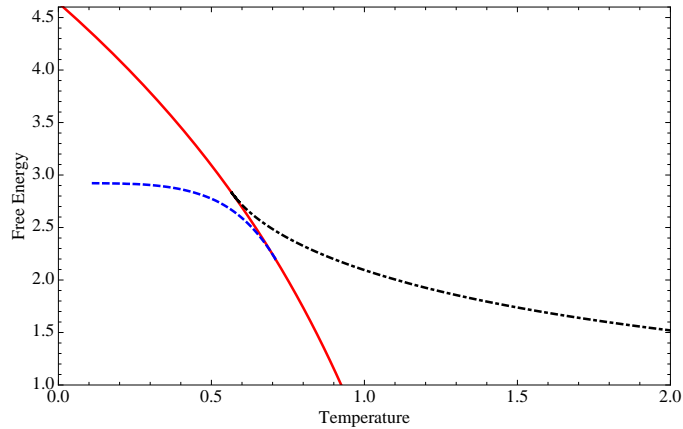} 
\caption{Free energies for the different black hole solutions. The dot-dashed black curve (lying on the $T>T_c$ side, $T_c\cong 0.56$) describes the free energy of 
the hairy black hole of sector I.
The red solid curve (i.e. the one shown in the interval $0<T<0.93$, but extending  to all $T$) 
represents the free energy of  the Reissner-Nordstr\" om black hole with $Q_1=Q_2=Q_3$. Finally, the dashed blue curve lying at $T<0.70$ describes the free energy of the sector II black hole.
}
\label{gubs}
\end{figure}

\newpage

\item {\it Condensation in Sector II}

The Lagrangian is given by (\ref{Gubs}). It describes a complex scalar with  $m^2 L^2=-3$ and $qL=\sqrt{3}$.
Comparing with the previous model, this scalar field has a greater mass and also greater charge. As discussed, increasing the mass lowers the
critical temperature, but increasing the charge raises it, so in this case it is not a priori  obvious whether the critical temperature
for this model will be lower or higher than the previous case.
We have carried out the explicit calculation, including backreaction.
We find that in this sector the critical temperature is $T_c\cong 0.70$, i.e. greater than $T_c\cong 0.56 $ of sector I.

The results are shown in Fig. \ref{fig3charge}b, which reproduces the results of \cite{Gubser:2009qm}.
Figure \ref{gubs} shows, in the same plot,  the free energy for the three different solutions: the Reissner-Nordstr\" om 
AdS solution, the sector I hairy black hole representing retrograde condensation, and the sector II hairy black hole providing
the dominant thermodynamics below $T_c\cong 0.70$.

In conclusion, among the two best candidates to condense  in the three equal charge ensemble, the
scalar mode of sector II is the one that condenses.
Thus the consistent truncation carried out in \cite{Gubser:2009qm} (particularized to the case when the Sasaki-Einstein manifold is $S^5$)
seems to indeed pinpoint the relevant
mode to study condensation in this ensemble. This is despite the fact that the scalar mode of sector II is not the one which is dual to the lowest dimension operator.
The scalar mode of sector I is dual to a operator of lower dimension, but it gives rise to hairy black holes at $T>T_c$ with 
a free energy that is higher than the free energy of the RN black hole representing the uncondensed phase.

\medskip

One important issue regards the stability of the $T=0$ limit of the hairy black hole of sector II.
In \cite{Bobev:2010de} it was pointed out that the 
the zero-temperature  solutions are domain walls interpolating between non-supersymmetric AdS solutions at the horizon
and AdS at infinity, which for compactifications based on spheres are unstable.
This fact may be interpreted in different ways. On one hand, it might indicate that, in reaching $T=0$,  there must be a quantum phase
transition to  a stable fixed point of the ${\cal N}=8$ potential. 
On the other hand, it could be that this sector just does not capture the dominant thermodynamic configuration even at finite temperature $T<T_c$, and that there is another sector within ${\cal N}=8$ supergravity containing black hole solutions that
 at sufficiently low temperatures dominate the thermodynamics and smoothly flow to a supersymmetric fixed point (which would thus ensure stability).

In an attempt to search for such a sector with a stable $T=0$ limit, we have examined  sectors that include the $SU(2)\times U(1)$ supersymmetric fixed point of ${\cal N}=8$ $SO(6)$ gauged supergravity
to see if it is possible to have a flow towards this point. The simplest consistent truncation of this type is achieved in sector II
with $A_1=A_2$ and a different $A_3\neq 0$, and setting $\td\eta_{2,3,4}=0$ and $\varphi_2=0$ in the sector II Lagrangian \eqref{sectorI},
with the charges given by (\ref{neoc}). The truncated Lagrangian is as follows
\bea
\cL &=& \sqrt{-g} \Big[ R - \tfrac12   (\p \varphi_1)^2 
-  \tfrac12 (\p \eta )^2  - \tfrac12 \sinh^2(\eta ) \big(\p \theta +  L^{-1} (2A_1-A_3)\big)^2  
\nn\\
&-& \tfrac12  e^{-{2\varphi_1\over\sqrt{6}}} F^1_{\mu\nu} F^{1,\ \mu\nu} 
 - \tfrac14  e^{{4\varphi_1\over\sqrt{6}}} F^3_{\mu\nu} F^{3,\ \mu\nu} 
- V \Big] \,,
\label{lsusy}
\eea
with
\be
V= \frac{e^{-\frac{4\varphi_1 }{\sqrt{6}}}}{L^2}  \cosh^2 (\tfrac12 {\eta})  \left( (1+2 e^{\sqrt{6}\varphi_1}) \cosh( \eta)  -8 e^{\frac{\sqrt{6}\varphi_1}{2}}-1   -6 e^{\sqrt{6}\varphi_1} \right)\ .
\label{susyV}
\ee

The resulting model indeed contains the  $SU(2)\times U(1)$ supersymmetric fixed point as a classical solution
of the equations and, moreover, one can write a consistent ansatz for  black holes with three equal charges (despite the  fact that $A_3\neq A_1=A_2$).
The numerical resolution of this system presents new complications. To the extent we were able to carry out the numerical analysis reliably, 
we found no  condensed phase in this sector, i.e.  no  black hole with three equal charges and $\td \eta_1$ hair with the required 
asymptotic behavior.

\end{itemize}

\section{Condensation from other sectors}

Since there are $42$ physical scalars in the five-dimensional ${\cal N}=8$ supergravity description,
a natural question is whether there could be a scalar not considered in our analysis that could condense earlier.
Clearly, an accurate answer to this point would require a long analysis. Nevertheless, one can guess which scalars could be  relevant thermodynamically from the following analysis:
 \begin{itemize}
 \item 
From the $({\bf 1},\tw)$, there are 2 neutral scalars (namely $\varphi_1$ and $\varphi_2$) and the three complex scalars that we have considered in sector I with  $U(1)\times U(1)\times U(1)$  charges $(\pm 2,0,0)$  and cyclic permutations.
The remaining scalars are 6 complex scalars  with charges $(\pm 1, \pm 1,0)$ (signs unrelated) and cyclic permutations. These last scalars have the same $m^2 L^2=-4$, but lower charges under each $U(1)$.
However, they could  compete with $\eta_{i}$ in the two-equal charge thermal ensemble, since they would have the same
charge under a diagonal $U(1)_D \subset U(1)\times U(1)$. In view of this possibility, we have explicitly investigated truncations including these charged scalar fields.

Starting from the symmetric scalar matrix $T_{mn}$, parametrising the $({\bf 1},\tw)$, one can perform truncations based on the following discrete $SO(6)$ transformations:
  \begin{align}
  \left( \begin{array}{cc} - \mathbb{I}_4 & \\ & +\mathbb{I}_2 \end{array} \right) \,, \qquad \left( \begin{array}{cccccc} & 1 &&&& \\ -1 &&&&& \\ &&& -1 && \\ && 1 &&& \\ &&&&& 1 \\ &&&& -1 & \end{array} \right) \,, \qquad
   \left( \begin{array}{ccccc} &&& 1 & \\ && -1 && \\  & - 1 &&& \\ 1 &&&& \\ &&&& \mathbb{I}_2 \end{array} \right) \,.
 \end{align}
This leads to a parametrisation of $T_{mn}$ that includes a dilaton $\varphi_1$ and a charged scalar field $\eta$ (in the gauge where the corresponding angular variable $\theta$ is set to zero) of the following form:
 \begin{align}
   \left( \begin{array}{cccccc} e^{\varphi_1} \cosh( \eta / \sqrt{2}) && e^{\varphi_1} \sinh( \eta / \sqrt{2})  &&& \\ & e^{\varphi_1} \cosh( \eta / \sqrt{2}) && - e^{\varphi_1} \sinh( \eta / \sqrt{2}) && \\ e^{\varphi_1} \sinh( \eta / \sqrt{2}) & & e^{\varphi_1} \cosh( \eta / \sqrt{2}) &&& \\ & - e^{\varphi_1} \sinh( \eta / \sqrt{2})&& e^{\varphi_1} \cosh( \eta / \sqrt{2}) && \\ &&&& e^{-2 \varphi_1} &  \\ &&&&& e^{-2 \varphi_1} \end{array} \right) \,.
 \end{align}
However, this leads to a Lagrangian that is identical to \eqref{tresnu} and \eqref{dosnu}, i.e.~these scalars not only have the same mass and charge, but also higher-order coefficients are the same. Therefore their thermodynamic properties will not differ from the truncation already considered in section 2.2.

\item
From the $({\bf 2},\tww)_+$, i.e. an anti-symmetric three-form, one finds
  the four complex scalars of sector II with charges $(\pm 1, \pm 1, \pm 1)$ (signs
unrelated) and in addition   two copies of three additional complex scalars with charges $(0,0,\pm 1)$ and cyclic
permutations. All these scalars have $m^2 L^2=-3$. Clearly, these six extra complex scalars have lower charges
in the three ensembles we have considered, so they are not expected to be relevant thermodynamically
(they are likely to condense at lower $T_c$ and probably with higher free energy, since this typically starts from 0 and becomes more negative as the temperature is lowered).
\end{itemize}
Therefore we expect to have captured the relevant scalar degrees of freedom for the various black hole ensembles. In particular, for the two-charge ensemble we do not find any other charged scalars that could condense.

\section{Concluding remarks}

In this paper we have examined the emergence of condensed phases in ${\cal N}=8$ supergravity originating from four different complex scalars in sector I and in sector II. We have found a new hairy black hole in the one-charge case (with the special property that at some $T_1 < T_c$ another phase transition seems to occur) and reproduced the known result of \cite{Gubser:2009qm} in the three-charge case. In addition, in both the two- and three-charge cases, hairy black holes were found in the unusual temperature range $T > T_c$. Interestingly, these hairy black holes are always subdominant in the free energy (which we referred to as retrograde condensation), while the opposite is the case for hairy black holes with $T < T_c$, in line with superconductivity {\it below} rather than {\it above} a certain critical temperature.


\medskip

The dual field theory of the present system is finite temperature ${\cal N}=4$ $SU(N)$ super Yang-Mills theory with three independent charge densities or chemical potentials.
The solutions we found for both sectors I and II are also solutions in the extended framework of the supersymmetric truncation of Section 2.1, and in the full ${\cal N}=8$ supergravity, since the  truncations are consistent.
Turning on scalars corresponds to adding deformations or having vacuum expectation values (depending on the asymptotic boundary conditions)
of their dual operators. 
The operators which are dual to a complex scalar mode of sector I, $Z_i=\eta_i e^{i\theta_i}$, $i=1,2,3$, with mass $m^2 L^2=-4 $, have conformal dimension defined by the standard 
AdS/CFT relation $\Delta (\Delta-4)=m^2L^2=-4$, i.e. $\Delta= 2$.
In addition, they must carry charge = 2 under the relevant $U(1)$. The natural candidate for the CFT operators which are dual to $Z_i$ are
\be
O_i ={\rm Tr}\big[\Phi_i^2\big]\ ,
\ee
where $\Phi_i$ are the three chiral superfields of ${\cal N}=4$ super Yang-Mills theory.
These are BPS operators.
The scalars of sector II, $\td Z_a=\td \eta_a e^{i\td \theta_a}$, are also dual to 
 BPS operators whose relevant component is  made with fermion bilinears \cite{Bobev:2010de}
\be
\td O_a ={\rm Tr}\big[\lambda_a \lambda_a\big]+{\rm h.c.}\ .
\ee
Our analysis suggests that in the three-charge ensemble, the first operator to condense is $\tilde O_4$, just as in the type IIB
truncations of  \cite{Gubser:2009qm} (for the case of $S^5$).
For the two charge ensemble, we have found no phase transition in the sectors we studied.
Finally, in the one-charge case we have found an intriguing condensation (corresponding to  a vacuum expectation value for 
$O_3$) where the hairy black hole rapidly looses its energy and reduces to a small size. We argued that this describes a new superconducting
phase with an order parameter that  becomes very large at some $T_1<T_c$. The specific heat has first a jump at $T_c$  --which is characteristic of second-order phase transitions-- and
then a very  high peak near $T_1$, where  entropy and internal energy
get suddenly  very small. 
It would be interesting to identify condensed matter systems with similar properties.

\section*{Acknowledgements}

We would like to thank Aristomenis Donos, Jerome Gauntlett, Maxim Mostovoy, Diego Rodriguez-Gomez and Paul Townsend for useful discussions.
F.A. is supported by a MEC FPU Grant No.AP2008-04553. D.R.~acknowledges support by a VIDI grant from the Netherlands Organisation for Scientific Research (NWO). J.R.~acknowledges support by MCYT Research
Grant No.  FPA 2007-66665 and Generalitat de Catalunya under project 2009SGR502.

\providecommand{\href}[2]{#2}\begingroup\raggedright\endgroup

\end{document}